\pdfoutput=1
\documentclass[a4paper,12pt]{article}

\usepackage{amsfonts}
\usepackage{mathrsfs}
\usepackage{amsmath}
\usepackage{amssymb}
\usepackage{framed}

\usepackage[medium]{titlesec}
\usepackage{bm}
\usepackage{cite}

\usepackage[normalem]{ulem}
\usepackage{extarrows}
\usepackage{slashed}
\usepackage{isodateo}
\usepackage{graphicx}
\usepackage[dvipsnames]{xcolor}
\usepackage[bookmarksnumbered=true,bookmarksopen=true]{hyperref}
 \hypersetup{colorlinks,%
             linkcolor=NavyBlue, %
             citecolor=PineGreen, %
             urlcolor=PineGreen}
\usepackage[hmargin=.7in,vmargin=1.1in]{geometry}
\usepackage{indentfirst}
\usepackage{booktabs}

\usepackage{bbm}

\newcommand{\FR}[2]{\displaystyle\frac{\,{#1}\,}{#2}}
\newcommand{\fr}[2]{\mbox{$\frac{\,{#1}\,}{#2}$}}
\newcommand{\n}{\nonumber}
\renewcommand{\rm}{\mathrm}

\graphicspath{{fig/}}

\def\bge{\begin{equation}}
\def\ede{\end{equation}}
\def\bga{\begin{aligned}}
\def\eda{\end{aligned}}
\def\bgb{\begin{bmatrix}}
\def\edb{\end{bmatrix}}
\def\bgp{\begin{pmatrix}}
\def\edp{\end{pmatrix}}
\def\bgm{\begin{matrix}}
\def\edm{\end{matrix}}
\def\bgs{\begin{subequations}}
\def\eds{\end{subequations}}

\def\di{{\mathrm{d}}}

\def\mb{\mathbf}

\def\pd{\partial}
\def\ld{{\mathscr{L}}}

\def\la{\langle}\def\ra{\rangle}

\def\to{\rightarrow}

\def\ii{\mathrm{i}}

\def\de{\delta}

\def\lam{\lambda}

\def\si{\sigma}

\def\aa{\mathsf{a}}
\def\bb{\mathsf{b}}
\def\cc{\mathsf{c}}

\def\2F1{{}_2\mathrm{F}_1}
\def\3F2{{}_3\mathrm{F}_2}

\usepackage{mdframed}

\newmdenv[skipabove=0pt,%
          skipbelow=5pt,%
          leftmargin=0pt,%
          rightmargin=0pt,%
          innertopmargin=-5pt,%
          innerbottommargin=7pt,%
          innerleftmargin=2pt,%
          innerrightmargin=2pt,%
          splittopskip=0pt,%
          splitbottomskip=0pt,%
          linewidth=0pt,%
          nobreak=true]%
          {keyeqn2}

\newmdenv[backgroundcolor=gray!15,%
          skipabove=0pt,%
          skipbelow=5pt,%
          leftmargin=0pt,%
          rightmargin=0pt,%
          innertopmargin=-5pt,%
          innerbottommargin=7pt,%
          innerleftmargin=2pt,%
          innerrightmargin=2pt,%
          splittopskip=0pt,%
          splitbottomskip=0pt,%
          linewidth=0pt,%
          nobreak=true]%
          {keyeqn}

\usepackage{titlesec}          
\titleformat{\section}
{\normalfont\fontsize{15}{20}\bfseries}{\thesection}{1em}{}

\newcommand{\wt}[1]{\mkern 2mu \widetilde{\mkern -2mu #1 \mkern -2mu}\mkern 2mu}
\newcommand{\wh}[1]{\mkern 2mu \widehat{\mkern-2mu#1\mkern-2mu}\mkern 2mu}

\newcommand{\fnemail}[1]{\footnote{Email: \href{mailto:#1}{\nolinkurl{#1}}}}

\begin{document}

\title{\Large\textbf{Cosmological Correlators at the Loop Level\\[2mm]}}

\author{Zhehan Qin$^{\,a,b\,}$\fnemail{qzh21@mails.tsinghua.edu.cn}\\[5mm]
$^a\,$\normalsize{\emph{Department of Physics, Tsinghua University, Beijing 100084, China} }\\
$^b\,$\normalsize{\emph{Department of Applied Mathematics and Theoretical Physics, University of Cambridge,}}\\\normalsize{\emph{Wilberforce Road, Cambridge, CB3 0WA, UK}}\\
}

\date{}
\maketitle

\vspace{20mm}

\begin{abstract}
\vspace{10mm}
Cosmological correlators encode rich information about physics at the Hubble scale and may exhibit characteristic oscillatory signals due to the exchange of massive particles. Although many 1-loop processes, especially those that break de Sitter (dS) boosts, can generate significant leading signals for various particle models in cosmological collider physics, the precise results for these correlators or their full signals remain unknown due to the lack of symmetry.
In this work, we apply the method of partial Mellin-Barnes (PMB) representation to the calculation of cosmological correlators at the loop level. As a first step, we use the PMB representation to calculate four-point cosmological correlators with bubble topology. We find that both the nonlocal and local signals arise from the factorized part, validating the cutting rules proposed in previous work, and are free from ultraviolet (UV) divergence. Furthermore, the UV divergence originates solely from the background piece and can be manifestly canceled by introducing the appropriate counterterm, similar to the procedure in flat spacetime. We also demonstrate how to renormalize the 1-loop correlators in Mellin space. After a consistency check with known results for the covariant case, we provide new analytical results for the signals generated from a nontrivial dS-boost-breaking bubble.
\end{abstract}

\newpage
\tableofcontents

\newpage
\section{Introduction}\label{sec_intro}
The $n$-point correlation functions at the end of inflation of massless primordial curvature fluctuations encode rich physics at the Hubble scale. These \emph{cosmological correlators}, also known as \emph{inflationary correlators}, establish the initial conditions for the subsequent Big Bang phase and provide the quantum sources responsible for the spatial inhomogeneity of our universe on large scales. Measuring these cosmological correlators through Cosmic Microwave Background (CMB) observations and Large Scale Structure (LSS) surveys presents a promising avenue for exploring high-energy physics at the inflationary scale.

In recent years, significant efforts have been dedicated to understanding the structure of these cosmological correlators \cite{Baumann:2022jpr,Maldacena:2011nz,Assassi:2012zq,Arkani-Hamed:2017fdk,Baumann:2017jvh,Arkani-Hamed:2018bjr,Arkani-Hamed:2018kmz,Baumann:2019oyu,Baumann:2020dch,Sleight:2019mgd,Sleight:2019hfp,Sleight:2020obc,Sleight:2021iix,Sleight:2021plv,Hillman:2019wgh,Pajer:2020wnj,Pajer:2020wxk,Goodhew:2020hob,Jazayeri:2021fvk,Melville:2021lst,Goodhew:2021oqg,Baumann:2021fxj,Gomez:2021qfd,Gomez:2021ujt,Bonifacio:2021azc,Meltzer:2021zin,Hogervorst:2021uvp,DiPietro:2021sjt,Cabass:2021fnw,Wang:2021qez,Premkumar:2021mlz,Hillman:2021bnk,Tong:2021wai,Heckelbacher:2022hbq,Pimentel:2022fsc,Wang:2022eop,Qin:2022lva,Jazayeri:2022kjy,Qin:2022fbv,Cabass:2022jda,Cabass:2022rhr,Cabass:2022oap,Xianyu:2022jwk,Bonifacio:2022vwa,Salcedo:2022aal,Lee:2022fgr,Qin:2023ejc,Werth:2023pfl,Pinol:2023oux,Qin:2023bjk,Qin:2023nhv,Lee:2023jby,Loparco:2023rug,AguiSalcedo:2023nds,De:2023xue,Stefanyszyn:2023qov,Xianyu:2023ytd,Green:2023ids,DuasoPueyo:2023kyh,Arkani-Hamed:2023bsv,Arkani-Hamed:2023kig,Chen:2023iix,Benincasa:2024leu,Benincasa:2024lxe,Werth:2024aui,Donath:2024utn,Du:2024hol,Fan:2024iek,Grimm:2024mbw,Melville:2024ove,Stefanyszyn:2024msm,Ema:2024hkj,Cohen:2024anu,Liu:2024xyi,He:2024olr,Baumann:2024ttn,Goodhew:2024eup,Green:2024fsz,Hang:2024xas,Baumann:2024mvm,Lee:2024sks,Werth:2024mjg,Chen:2024glu,Gasparotto:2024bku,De:2024zic}.
In particular, heavy particles can be spontaneously produced from the vacuum due to the breaking of time translation symmetry and the lack of energy conservation. Through their interactions with curvature fluctuations, these particles leave characteristic imprints on the cosmological correlators, manifesting as oscillations on the logarithmic scale in momentum space. Such oscillations can be utilized to extract physical information about these heavy particles (with masses comparable to the Hubble parameter during inflation), including their mass, spin, and types of interactions. This paradigm, first explored in the context of quasi-single field inflation \cite{Chen:2009zp,Baumann:2011nk,Noumi:2012vr}, has been termed Cosmological Collider (CC) physics \cite{Arkani-Hamed:2015bza}.
In this framework, a substantial amount of phenomenological work has been conducted\cite{Chen:2009we,Chen:2009zp,Baumann:2011nk,Chen:2012ge,Pi:2012gf,Noumi:2012vr,Gong:2013sma,Arkani-Hamed:2015bza,Chen:2015lza,Chen:2016nrs,Chen:2016uwp,Chen:2016hrz,Lee:2016vti,Chen:2017ryl,An:2017hlx,An:2017rwo,Iyer:2017qzw,Kumar:2017ecc,Tong:2018tqf,Chen:2018sce,Saito:2018omt,Chen:2018xck,Chen:2018cgg,Chua:2018dqh,Kumar:2018jxz,Wu:2018lmx,Li:2019ves,Alexander:2019vtb,Lu:2019tjj,Hook:2019zxa,Hook:2019vcn,Kumar:2019ebj,Liu:2019fag,Wang:2019gbi,Wang:2019gok,Wang:2020uic,Li:2020xwr,Wang:2020ioa,Fan:2020xgh,Bodas:2020yho,Aoki:2020zbj,Maru:2021ezc,Lu:2021gso,Sou:2021juh,Lu:2021wxu,Pinol:2021aun,Cui:2021iie,Tong:2022cdz,Reece:2022soh,Chen:2022vzh,Maru:2022bhr,Niu:2022quw,Niu:2022fki,Aoki:2023tjm,Chen:2023txq,Tong:2023krn,Jazayeri:2023xcj,Jazayeri:2023kji,Chakraborty:2023qbp,Chakraborty:2023eoq,Aoki:2023wdc,Craig:2024qgy,McCulloch:2024hiz,Wu:2024wti,Aoki:2024uyi,Aoki:2024jha,Bodas:2024hih}, and people are using real cosmological data to search and constrain such oscillatory CC signals \cite{Cabass:2024wob,Sohn:2024xzd}.

However, calculating these correlators and/or their CC signals in curved spacetime is extremely challenging compared to flat spacetime. The need for accurate templates for cosmological observations, as well as a deeper understanding of quantum field theories (QFTs) in de Sitter (dS) space, has led to the development of many analytical methods. These methods include, but are not limited to: the cosmological bootstrap from a boundary perspective \cite{Arkani-Hamed:2018kmz,Baumann:2019oyu,Baumann:2020dch,Pimentel:2022fsc,Jazayeri:2022kjy,Qin:2022fbv,Qin:2023ejc,Aoki:2024uyi,Gasparotto:2024bku} which solves differential equations derived from boundary conformal symmetry and/or the equations of motion that the cosmological correlators satisfy; and the Mellin-Barnes representation \cite{Sleight:2019mgd,Sleight:2019hfp,Sleight:2020obc}, which leverages the dilation symmetry of dS. Based on this, the method of partial Mellin-Barnes (PMB) representation \cite{Qin:2022lva,Qin:2022fbv} has been developed and has proven to be very suitable and useful for dS calculations \cite{Qin:2023bjk,Qin:2023nhv,Xianyu:2023ytd,Qin:Local}.
Other methods include spectral decomposition \cite{Xianyu:2022jwk} using the Källén-Lehmann representation \cite{Marolf:2010zp,Loparco:2023rug}, the dispersion integral \cite{Liu:2024xyi} based on the nonanalyticity of the correlators, and an off-shell approach \cite{Werth:2024mjg} that relies on the spectral representation of bulk field propagators \cite{Sleight:2020obc,Melville:2024ove}. Additionally, see \cite{Wang:2021qez,Werth:2023pfl,Pinol:2023oux,Werth:2024aui} for numerical techniques, and the integration by parts (IBP) technique \cite{Chen:2023iix,Chen:2024glu} that shows its potential for numerical calculations.

Most of the above methods apply to tree-level processes and have provided significant insights into quantum processes in dS, but loops are also important. For instance, the leading CC signals generated by fermions or charged particles come from loop-level processes. However,
compared to the impressive results at the tree level, our understanding of loop diagrams remains quite limited. The only results we have for loop-level correlators containing CC signals are the covariant bubble diagrams,\footnote{By covariant bubble, we mean that the bulk two-point correlator $\la \si_1\si_2(x)  \si_1\si_2(y)\ra$ in the bubble respects all dS isometries. A counterexample is $\la \si'^2(x)  \si'^2(y)\ra$ that breaks dS boosts, which we will consider in Sec.\ \ref{sec_break}.} where the calculation, using either spectral decomposition \cite{Xianyu:2022jwk} or dispersion integral \cite{Liu:2024xyi}, heavily relies on the full dS isometry of the bulk propagators.\footnote{The dispersion integral itself does not rely on the dS isometries. However, we need the result of the signal part of the full correlator (which generates the nonanalyticity) from spectral decomposition as an input.} On the other hand, CC signals suffer from Boltzmann suppression and are exponentially suppressed by the heavy mass of exchanged particles, and signals generated from loop diagrams are suppressed twice, making them unlikely to be observed. Therefore, people are building models with dS-boost breaking to overcome the Boltzmann suppression and produce large signals at the loop level, which are of greater interest from a phenomenological perspective. See e.g. \cite{Chen:2018xck,Lu:2019tjj,Wang:2019gbi,Wang:2020ioa,Cui:2021iie,Tong:2022cdz}. However, the analytical results for these important processes remain unknown, though we have the partial result for the nonlocal signals using the PMB representation \cite{Qin:2022lva}.

In this work, we continue our exploration of the application of the PMB representation to the calculation of cosmological correlators. Having successfully applied it to arbitrary tree-level processes \cite{Qin:2022fbv, Xianyu:2023ytd}, we now initiate a bulk calculation of cosmological correlators involving 1-loop massive exchanges through the PMB representation. We will detail the computation of the multi-layered Mellin integral and demonstrate that our results are consistent with known outcomes for the dS-covariant case. Since the PMB representation relies solely on dilation symmetry, it is particularly well-suited for dS-boost-breaking models that can generate large CC signals. As a first step, we consider a simple but nontrivial example with dS-boost breaking and present the explicit analytical results for the signals using PMB representation.

\paragraph{Outline of this work}
The rest of the paper is organized as follows.
 In Sec.\ \ref{sec_covariant}, we briefly review the method of PMB representation.
 We also recall the covariant loop seed integral \eqref{eq_loopseed} originally defined in \cite{Xianyu:2022jwk} that can be used to express many 1-loop exchanges with dS covariant interaction for the massive particles, and we generalize it to arbitrary internal masses but we fix in $3+1$ dimension. We apply PMB representation to this loop seed integral and complete the time and loop momentum integrals. We are left with multi-layered Mellin integrals \eqref{eq_JPMB}-\eqref{eq_BGPMB} that wait to be computed.

We then compute these Mellin integrals in Sec.\ \ref{sec_calculation}. After carefully identifying the pole structure of the integrand, we can complete the Mellin integral using the residue theorem by selecting residues at the appropriate series of poles. We carry out the calculation of the factorized and the time-ordered part of the loop seed integral in Sec.\ \ref{sec_JF} and Sec.\ \ref{sec_JTO}, respectively. We find that the factorized part of the integral is free from ultraviolet (UV) divergence, and all the CC signals originate solely from this factorized part, though it could also contain a nonvanishing background piece. On the other hand, the time-ordered part contributes to the background piece that exhibits UV divergence, and the divergent part manifests as a contribution from a contact graph, but with a divergent coefficient. This means we can renormalize the background (and thus the loop seed integral or the correlator) by adding a counterterm to the Lagrangian, similar to the case in flat spacetime. We summarize our results in Sec.\ \ref{sec_result}. We check the signal part is consistent with known results for the equal-mass case, and we also present new CC signals from different-mass bubbles.

In Sec.\ \ref{sec_break}, we apply the PMB representation to a toy model with nontrivial dS-boost breaking. Following the steps in Sec.\ \ref{sec_calculation}, we can easily compute the CC signal of this model, and we provide the full analytical result for the signals, which cannot be obtained from any other methods to the best of our knowledge. We also discuss the signals' behavior in the (hierarchical) squeezed limit, and compare them with signals from the covariant bubble. The conclusion and outlooks are given in Sec.\ \ref{sec_conclusion}.

There are three appendices following the main text. We collect our notations of frequently used special functions and some of their properties in App.\ \ref{app_A}.
Useful integrals formulae are listed in App.\ \ref{app_B}. We put the further details of our calculations in App.\ \ref{app_C}.

\paragraph{Notation and convention}
We work in the slow-roll limit of the inflation and the spacetime metric reads $\di s^2 = a^2(\tau) (-\di\tau^2+\di\mb x^2)$, where $\tau\in(-\infty,0)$ is the conformal time, and $\mb x$ is the spatial comoving coordinate. The scale factor $a(\tau)\equiv -1/(H\tau)$, where $H$ is the constant Hubble parameter. Throughout this work, we set $H\equiv 1$.

\begin{figure}
\centering 
\includegraphics[width=0.6\textwidth]{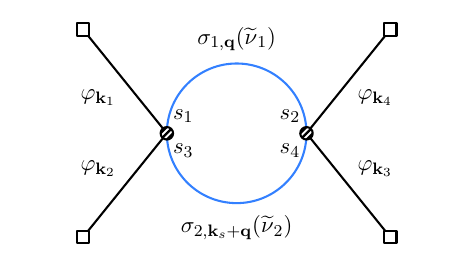} 
\caption{The four-point function of inflaton fluctuations $\varphi$ mediated by a bubble of massive scalars $\si_1$ and $\si_2$ with arbitrary masses $\wt\nu_1$ and $\wt\nu_2$. We also use this diagram to represent other processes with the same bubble topology: The external lines can be replaced by the conformal scalar $\phi$, the internal lines can be identical $\wt\nu_1=\wt\nu_2=\wt\nu$ (where an extra symmetry factor $1/2$ should be added), and we do not specify the interactions at the two vertices. We use PMB representation to calculate this diagram, and we always label the four introduced Mellin variables $s_i$ ($i=1,2,3,4$) as shown in the plot.}
  \label{fig_bubble}
\end{figure}

The 1-loop diagram with bubble topology shown in Fig.\ \ref{fig_bubble} is the main target of this work, which is determined by the four external spatial momenta $\mb k_i$
$(i = 1, 2, 3, 4)$, and the $s$-channel momentum is defined as $\mb k_s \equiv \mb k_1 + \mb k_2$. We use the corresponding italic letter to denote the magnitude of a three-momentum, e.g. $k_i\equiv |\mb k_i|$ and $k_s\equiv |\mb k_s|$.
We shall frequently use the momentum ratios $r_1 \equiv k_s/k_{12}$ and $r_2 \equiv  k_s/k_{34}$, where we use a shorthand notation for sums of several indexed quantities such as
$k_{12} \equiv k_1 + k_2$.
Other similar shorthands include $\bar p_{12} = p_1-p_2$, etc.

\section{Covariant Loop Seed Integral under PMB}
\label{sec_covariant}
In this section, we will briefly review the PMB representation method introduced in \cite{Qin:2022lva,Qin:2022fbv}. We will formulate the covariant loop seed integral, which can represent various 1-loop processes with dS-covariant interactions in the bulk, as a multi-layer Mellin integral. This formulation is prepared for computation in Sec.\ \ref{sec_calculation}.
\subsection{PMB revisited}
\label{sec_PMB}
Mellin-Barnes (MB) representation is an integral representation with the eigenfunctions of dilation symmetry (i.e., monomials) as the integral kernel, which is the dS counterpart of the Fourier transform in flat spacetime.
With the MB representation, we can resolve complicated special functions appearing in the mode functions of fields into simple power and exponential functions, which simplifies the time and loop momentum integrals significantly.
In particular, the mode function of a scalar in the principal series with mass $m>3/2$ is given by:
\bge
u(k,\tau) = \FR{\sqrt\pi}2e^{-\pi\wt\nu/2}(-\tau)^{3/2}\rm H_{\ii\wt\nu}^{(1)}(-k\tau),
\ede
where $\wt\nu\equiv \sqrt{m^2-9/4}$ is the (positive) real parameter characterizing the mass of the heavy field.\footnote{Throughout this work we consider scalars in the principal series that can generate oscillatory CC signals, though all the results can be analytically continued to the complementary series if the theory does not diverge in the infrared (IR). The only difference is that $\wt\nu$ will be pure imaginary, $\wt\nu \to \pm\ii\sqrt{9/4-m^2}$.}

We will use the diagrammatic rule \cite{Chen:2017ryl} of Schwinger-Keldysh (SK) formalism \cite{Schwinger:1960qe,Feynman:1963fq,Keldysh:1964ud,Weinberg:2005vy} to calculate the cosmological correlators, where the bulk-to-bulk propagators read:
\begin{align}
\label{eq_D-+}
&D_{-+}^{(\wt\nu)}(k;\tau_1,\tau_2) = u(k,\tau_1)u^*(k,\tau_2),\\
\label{eq_D+-}
&D_{+-}^{(\wt\nu)}(k;\tau_1,\tau_2) = [D_{+-}^{(\wt\nu)}(k;\tau_1,\tau_2)]^*,\\
\label{eq_D++}
&D_{\pm\pm}^{(\wt\nu)}(k;\tau_1,\tau_2) = D_{\mp\pm}^{(\wt\nu)}(k;\tau_1,\tau_2) \theta(\tau_1-\tau_2) + D_{\pm\mp}^{(\wt\nu)}(k;\tau_1,\tau_2)\theta(\tau_2-\tau_1).
\end{align}
To apply the PMB representation,
we make use of the following inverse Mellin transforms of Hankel functions \cite{nist:dlmf}:
\begin{align}
&\rm H_{\ii\wt\nu}^{(1)}(-k\tau) = \FR{-\ii e^{\pi\wt\nu/2}}{\pi} \int_{-\ii\infty}^{+\ii\infty} \FR{\di s}{2\pi\ii}\,e^{\ii\pi s}\Big(\FR{-k\tau}2\Big)^{-2s}
\Gamma\Big[s+\FR{\ii\wt\nu}2,s-\FR{\ii\wt\nu}2 \Big],\\
&\rm H_{-\ii\wt\nu}^{(2)}(-k\tau) = \FR{\ii e^{\pi\wt\nu/2}}{\pi} \int_{-\ii\infty}^{+\ii\infty} \FR{\di s}{2\pi\ii}\,e^{-\ii\pi s}\Big(\FR{-k\tau}2\Big)^{-2s}
\Gamma\Big[s+\FR{\ii\wt\nu}2,s-\FR{\ii\wt\nu}2 \Big].
\end{align}
Plugging these into the bulk-to-bulk propagators of the massive scalar, we obtain their PMB representation:\footnote{Here we follow our notations in \cite{Qin:Local}, and we recommend readers refer to \cite{Qin:Local} for the PMB representation of arbitrary graphs.}
\begin{align}
\label{eq_DPMB}
D_{\aa\bb}^{(\wt\nu)}(k;\tau_1,\tau_2) =& \int_{-\ii\infty}^{+\ii\infty} \FR{\di s_1}{2\pi\ii}\FR{\di s_2}{2\pi\ii}\, N_{\aa\bb}(s_1,s_2;\tau_1,\tau_2)(-\tau_1)^{3/2-2s_1}(-\tau_2)^{3/2-2s_2}k^{-2s_{12}}\n\\
&\times \Gamma\Big[s_1+\FR{\ii\wt\nu}2,s_1-\FR{\ii\wt\nu}2,s_2+\FR{\ii\wt\nu}2,s_2-\FR{\ii\wt\nu}2\Big].
\end{align}
Here we have introduced the nesting function $N_{\aa\bb}$, which is a constant (regarding $\tau_{1,2}$) for the factorized opposite-sign propagators $D^{(\wt\nu)}_{\pm\mp}$ \eqref{eq_D-+} and \eqref{eq_D+-}, and a combination of Heaviside $\theta$ functions for the nested same-sign propagators $D^{(\wt\nu)}_{\pm\pm}$ \eqref{eq_D++}:
\begin{align}
\label{eq_Npmmp}
N_{\pm\mp}(s_1,s_2;\tau_1,\tau_2) \equiv&~ N_{\pm\mp}(s_1,s_2) =  2^{2s_{12}-2}\pi^{-1}e^{\mp\ii\pi(s_1-s_2)},\\
\label{eq_Npmpm}
N_{\pm\pm}(s_1,s_2;\tau_1,\tau_2) =&~  N_{\mp\pm}(s_1,s_2)\theta(\tau_1-\tau_2) + N_{\pm\mp}(s_1,s_2)\theta(\tau_2-\tau_1).
\end{align}

As we can see, the bulk propagators are expressed as a double-layered Mellin integral, where the integrand consists solely of simple functions (power and theta) of time and momentum. Furthermore, the dependencies on time and momentum are factorized, allowing us to analyze or even analytically compute the time and loop momentum integrals separately.

\subsection{Loop seed integral}
\label{sec_loopseed}
The main target we will calculate in this section is the covariant \emph{loop seed integral}, defined as:\footnote{Notice that definition of the loop seed integral here is slightly different from \cite{Xianyu:2022jwk}: First, we do not need dimensional regularization to regulate the UV divergence, thus we fix the spatial dimension to be $d=3$; Second, we leave the masses of internal lines in the bubble arbitrary as $\wt\nu_1$ and $\wt\nu_2$, and thus we do not have the symmetry factor $1/2$.}
\begin{align}
\label{eq_loopseed}
\mathcal J^{p_1p_2}_{\wt\nu_1,\wt\nu_2}(k_{12},k_{34};k_s) \equiv & \sum_{\aa,\bb=\pm}(-\aa\bb) k_s^{5+p_{12}}\int_{-\infty}^0\di\tau_1\di\tau_2\,(-\tau_1)^{p_1}(-\tau_2)^{p_2}e^{\aa\ii k_{12}+\bb\ii k_{34}}\n\\
&\times \int \FR{\di^3\mb q}{(2\pi)^3}\, D_{\aa\bb}^{(\wt\nu_1)}(q;\tau_1,\tau_2)D_{\aa\bb}^{(\wt\nu_2)}(|\mb k_s + \mb q|;\tau_1,\tau_2).
\end{align}
Here we assume the integral is IR-safe (see below for further explanations)
 and the factor $k_s^{5+p_{12}}$ is introduced to make the loop seed integral dimensionless.
Notably, the loop seed integral \eqref{eq_loopseed} is scale-invariant, which means it depends on only two independent variables. For convenience, we define the following two dimensionless ratios:
\bge
\label{eq_ratio}
r_1 \equiv \FR{k_s}{k_{12}},\qquad r_2\equiv \FR{k_s}{k_{34}}.
\ede
Then the loop seed integral \eqref{eq_loopseed} can be regarded as a function of these two ratios, namely:
\bge
\wt{\mathcal J}^{p_1p_2}_{\wt\nu_1,\wt\nu_2}(r_1,r_2) \equiv \mathcal J^{p_1p_2}_{\wt\nu_1,\wt\nu_2}(k_{12},k_{34};k_s).
\ede
Unless specifically emphasized, we will omit the superscript tilde and also the subscript ${}_{\wt\nu_1,\wt\nu_2}$ in the rest of this paper just for conciseness, and we will use $\mathcal J^{p_1p_2}(r_1,r_2)$ and $\mathcal J^{p_1p_2}(k_{12},k_{34};k_s)$ changeably.

As illustrated in \cite{Xianyu:2022jwk}, many one-loop correlators with covariant bubble exchange can be expressed via this loop seed integral \eqref{eq_loopseed}. For instance, consider a bubble diagram with the following general interaction that respects the spatial translation and rotation symmetry (where we omit all the coupling constants):
\bge
\label{eq_conformalL}
\Delta \ld = -\FR12 a^4(-\tau_1)^{2+p_1}\phi^2\si_1\si_2 -\FR12 a^4(-\tau_1)^{2+p_2}\phi^2\si_1\si_2.
\ede
Here $\phi$ denotes the conformal scalar with bulk-to-boundary propagators:
\bge
C_\pm(k;\tau) = \FR{\tau\tau_0}{2k}e^{\pm\ii k\tau},
\ede
and $\tau_0$ is the IR cutoff, and $\si_{1,2}$ are two massive scalar with mass parameters $\wt\nu_{1,2}$. Following the diagrammatic rule\cite{Chen:2017ryl}, the $s$-channel four-point correlator of conformal scalar contributed from Fig.\ \ref{fig_bubble} is:\footnote{Here we assume the left and right vertices in Fig.\ \ref{fig_bubble} are generated from the first and second terms in \eqref{eq_conformalL}, respectively. In principle the $s$-channel correlator also receive contributions from $\mathcal J^{p_2p_1}_{\wt\nu_1,\wt\nu_2}(r_1,r_2)$, $\mathcal J^{p_1p_1}_{\wt\nu_1,\wt\nu_2}(r_1,r_2)$, $\mathcal J^{p_2p_2}_{\wt\nu_1,\wt\nu_2}(r_1,r_2)$ if $p_1\neq p_2$.}
\bge
\la \phi_{\mb k_1}\phi_{\mb k_2}\phi_{\mb k_3}\phi_{\mb k_4} \ra'_s
\supset \FR{\tau_0^4}{16k_1k_2k_3k_4k_s^{5+p_{12}}} \mathcal J^{p_1p_2}_{\wt\nu_1,\wt\nu_2}(r_1,r_2),
\ede 
where a prime means the momentum conservation factor $(2\pi)^3\de(\mb k_1+\mb k_2+\mb k_3+\mb k_4)$ has been dropped out. If we further assume dilation symmetry in the Lagrangian \eqref{eq_conformalL}, we will have $p_1=p_2=-2$, and thus:
\bge
\label{eq_conformal4pt}
\la \phi_{\mb k_1}\phi_{\mb k_2}\phi_{\mb k_3}\phi_{\mb k_4} \ra'_s
= \FR{\tau_0^4}{16k_1k_2k_3k_4k_s}\mathcal J^{-2,-2}_{\wt\nu_1,\wt\nu_2}(r_1,r_2).
\ede

Similarly, if we consider the correlation function of massless inflaton $\varphi$ whose bulk-to-boundary propagators are given by:
\bge
G_\pm(k;\tau) = \FR{1\mp\ii k\tau}{2k^3}e^{\pm\ii k\tau},
\ede
with the general interaction:
\bge
\label{eq_masslessL}
\Delta\ld  = -\FR12 a^2(-\tau)^{p_1}\varphi'^2\si_1\si_2 -\FR12 a^2(-\tau)^{p_2}\varphi'^2\si_1\si_2,
\ede
where $\varphi$ is derivatively coupled to avoid IR divergence, then the corresponding $s$-channel four-point correlator is:
\bge
\label{eq_massless4ptgeneral}
\la \varphi_{\mb k_1}\varphi_{\mb k_2}\varphi_{\mb k_3}\varphi_{\mb k_4} \ra'_s
\supset \FR{1}{16k_1k_2k_3k_4k_s^{5+p_{12}}} \mathcal J^{p_1p_2}_{\wt\nu_1,\wt\nu_2}(r_1,r_2).
\ede
If we assume dilation symmetry in the Lagrangian, then $p_1=p_2=0$ and thus:
\bge
\label{eq_massless4pt}
\la \varphi_{\mb k_1}\varphi_{\mb k_2}\varphi_{\mb k_3}\varphi_{\mb k_4} \ra'_s
= \FR{1}{16k_1k_2k_3k_4k_s^5} \mathcal J^{0,0}_{\wt\nu_1,\wt\nu_2}(r_1,r_2).
\ede
Notice that although the Lagrangian \eqref{eq_masslessL} breaks both dilation and dS boosts for arbitrary values of $p_{1,2}$, the bulk two-point correlator $\la\si_1\si_2(x)\si_1\si_2(y)\ra$ in the bubble (see Fig.\ \ref{fig_bubble}) respects all dS isometries and thus enables a spectral decomposition. In this sense, we still refer to it as a covariant bubble.

The loop seed integral \eqref{eq_loopseed} can also express various bispectra and power spectra generated from bubble exchange, with one or both ratios $r_{1,2}$ sent to the folded limit $1$. We recommend readers refer to \cite{Xianyu:2022jwk} for more examples.

One thing to mention is that the assumption of IR safety requires $p_{1,2}$ to be bounded from below.  For most cases we are concerned, $p_{1,2}$ are integers and $p_{1,2}\geq -2$, as the two examples \eqref{eq_conformal4pt} and \eqref{eq_massless4pt} shown above.\footnote{Though there are also models involving non-integer $p$'s, see e.g. \cite{Bodas:2020yho,Chen:2022vzh}.} 
We can verify that the loop seed integral \eqref{eq_loopseed} is convergent in the IR when 
$p_{1,2}\geq -2$ by performing a late-time expansion of the integrand and power counting. However, determining the lower bound that saturates the IR convergence is nontrivial, and we hope to address this in future work.

The covariant loop seed integral \eqref{eq_loopseed} has been calculated in \cite{Xianyu:2022jwk} via spectral decomposition, and we will calculate it using PMB representation in Sec.\ \ref{sec_calculation} as a first trial.
Applying the PMB representation \eqref{eq_DPMB}, we can rewrite the loop seed integral \eqref{eq_loopseed} in the Mellin space:
\bge
\label{eq_loopPMB}
\mathcal J^{p_1p_2}(r_1,r_2) = k_s^{5+p_{12}} \int_{-\ii\infty}^{+\ii\infty} \Big[\prod_{i=1}^4\FR{\di s_i}{2\pi\ii}\Big]\, \mathcal T(k_{12},k_{34}) \times \mathcal L(k_s)\times \wt\Gamma(s_1,s_2;s_3,s_4),
\ede
where $\wt\Gamma(s_1,s_2;s_3,s_4)$ denotes the product of $\Gamma$ functions in the PMB representation of the propagators \eqref{eq_DPMB}:
\bge
\label{eq_Gamma}
\wt\Gamma(s_1,s_2;s_3,s_4)\equiv \prod_{i=1,2}\Gamma\Big[ s_i+\FR{\ii\wt\nu_1}2,s_i-\FR{\ii\wt\nu_1}2\Big]\times \prod_{j=3,4} \Gamma\Big[ s_j+\FR{\ii\wt\nu_2}2,s_j-\FR{\ii\wt\nu_2}2\Big],
\ede
while $\mathcal T$ is the time integral, which is a function of external energies $k_{12}$ and $k_{34}$:
\begin{align}
\label{eq_T}
\mathcal T(k_{12},k_{34}) =&~ \sum_{\aa,\bb=\pm}(-\aa\bb) \int_{-\infty}^0 \di\tau_1\di\tau_2\,(-\tau_1)^{3+p_1-2s_{13}}(-\tau_2)^{3+p_2-2s_{24}}e^{\aa\ii k_{12}\tau_1+\bb\ii k_{34}\tau_2}\n\\
&\times N_{\aa\bb}(s_1,s_2;\tau_1,\tau_2)N_{\aa\bb}(s_3,s_4;\tau_1,\tau_2),
\end{align}
and $\mathcal L$ is the loop integral as a function of exchanged momentum $k_s$:
\bge
\label{eq_L}
\mathcal L(k_s) = \int \FR{\di^3\mb q}{(2\pi)^3}\,|\mb q|^{-2s_{12}}|\mb k_s + \mb q|^{-2s_{34}}.
\ede

Thanks to the simplicity of integrands under PMB representation, both the time and loop integrals can be computed analytically. The loop integral \eqref{eq_L} is straightforward:
\bge
\label{eq_Lresult}
\mathcal L(k_s) = \FR{k_s^{3-2s_{1234}}}{(4\pi)^{3/2}}\Gamma\left[\bgm
s_{1234}-\fr32,\fr32-s_{12},\fr32-s_{34}\\ 3-s_{1234},s_{12},s_{34}
\edm\right],
\ede
while the time integral \eqref{eq_T} is more involved due to the time orderings in the product of nesting functions that cannot be removed. However, no matter how complicated the product is, the nesting function involving two times $\tau_{1,2}$ can always be decomposed and expressed as a combination of two independent possibilities: Either $\tau_1$ and $\tau_2$ are factorized so that the nesting function is a constant, or they are chronal and thus the nesting function is (proportional to) $\theta(\tau_2-\tau_1)$. Notice that the achronal case can be rewritten into this basis using the identity $\theta(\tau_1-\tau_2) = 1-\theta(\tau_2-\tau_1)$.

Without loss of generality, we assume $k_{12}>k_{34}$. For later convenience, we can decompose the product of the same-sign nesting functions into a factorized part and a time-ordered (chronal) part as the following:
\begin{align}
\label{eq_Nsimp}
&~N_{\pm\pm}(s_1,s_2;\tau_1,\tau_2)N_{\pm\pm}(s_3,s_4;\tau_1,\tau_2)\n\\
=&~ \Big[ N_{\mp\pm}(s_1,s_2)\theta(\tau_1-\tau_2) + N_{\pm\mp}(s_1,s_2)\theta(\tau_2-\tau_1)\Big]
\Big[ N_{\mp\pm}(s_3,s_4)\theta(\tau_1-\tau_2) + N_{\pm\mp}(s_3,s_4)\theta(\tau_2-\tau_1)\Big]\n\\
= &~ N_{\mp\pm}(s_1,s_2)N_{\mp\pm}(s_3,s_4) + \Big[ N_{\pm\mp}(s_1,s_2)N_{\pm\mp}(s_3,s_4)-N_{\mp\pm}(s_1,s_2)N_{\mp\pm}(s_3,s_4)\Big]\theta(\tau_2-\tau_1),
\end{align}
while the opposite-sign nesting functions are always factorized:
\bge
\label{eq_NFAC}
N_{\pm\mp}(s_1,s_2;\tau_1,\tau_2)N_{\pm\mp}(s_3,s_4;\tau_1,\tau_2) = N_{\pm\mp}(s_1,s_2)N_{\pm\mp}(s_3,s_4).
\ede
Then we split the time integral $\mathcal T$ into a factorized part $\mathcal T_{\text{F},>}$ and a time-ordered part $\mathcal T_{\text{TO},>}$ according to the decomposition of the product of nesting functions \eqref{eq_Nsimp} and \eqref{eq_NFAC}. Using the formulae \eqref{eq_1site} and \eqref{eq_2site}, we can complete the time integrals (see App.\ \ref{app_B}):
\begin{align}
\label{eq_Tresult}
\mathcal T(k_{12},k_{34})  =&~ \mathcal T_{\text{F},>}(k_{12},k_{34})+ \mathcal T_{\text{TO},>}(k_{12},k_{34}),\\
\label{eq_TF}
\mathcal T_{\text{F},>}(k_{12},k_{34}) =&~ \FR{2^{2s_{1234}}}{8\pi^2}\bigg[\cos \FR{\pi \bar p_{12}}2 -\cos\Big(2\pi s_{13} - \FR{\pi p_{12}}2\Big)\bigg] k_{12}^{-4-p_1+2s_{13}} k_{34}^{-4-p_2+2s_{24}}\n\\&\times \Gamma\Big[ 4+p_1-2s_{13},4+p_2-2s_{24}\Big],\\
\label{eq_TTO}
\mathcal T_{\text{TO},>}(k_{12},k_{34})=&~ \FR{2^{2s_{1234}}}{4\pi^2} \sin[\pi(s_{13}-s_{24})]\sin\Big( \FR{\pi p_{12}}2 - \pi s_{1234}\Big) k_{12}^{-8-p_{12}+2s_{1234}}\n\\&\times {}_2\mathcal F_1\left[\bgm4+p_2-2s_{24},8+p_{12}-2s_{1234}\\5+p_2-2s_{24}\edm\middle| -\FR{k_{34}}{k_{12}}\right].
\end{align}
Here ${}_2\mathcal F_1$ denotes the dressed hypergeometric function, see App.\ \ref{app_A}.

After plugging the results of the time integral \eqref{eq_Tresult}-\eqref{eq_TTO} and also the loop integral \eqref{eq_Lresult} back into the PMB representation \eqref{eq_loopPMB}, we obtain the decomposition for the loop seed integral \eqref{eq_loopseed}:
\begin{align}
\label{eq_JPMB}
\mathcal J^{p_1p_2}(r_1,r_2)  =&~ \mathcal J_{\text{F},>}^{p_1p_2}(r_1,r_2)+ \mathcal J^{p_1p_2}_{\text{TO},>}(r_1,r_2),\\
\label{eq_SPMB}
\mathcal J^{p_1p_2}_{\text{F},>}(r_1,r_2) =&~ k_s^{5+p_{12}} \int_{-\ii\infty}^{+\ii\infty} \Big[\prod_{i=1}^4\FR{\di s_i}{2\pi\ii}\Big]\, \mathcal T_{\text{F},>}(k_{12},k_{34}) \times \mathcal L(k_s)
\times \wt\Gamma(s_1,s_2;s_3,s_4),\\
\label{eq_BGPMB}
\mathcal J^{p_1p_2}_{\text{TO},>}(r_1,r_2) =&~ k_s^{5+p_{12}} \int_{-\ii\infty}^{+\ii\infty} \Big[\prod_{i=1}^4\FR{\di s_i}{2\pi\ii}\Big]\, \mathcal T_{\text{TO},>}(k_{12},k_{34}) \times \mathcal L(k_s)
\times \wt\Gamma(s_1,s_2;s_3,s_4).
\end{align}

It is worth mentioning that the above decomposition \eqref{eq_JPMB} is aligned with the cutting rules \cite{Tong:2021wai,Qin:2023bjk,Qin:2023nhv,Ema:2024hkj,Qin:Local} for the cosmological collider signals.\footnote{Cutting rules on wavefunction coefficients are discussed in e.g. \cite{Melville:2021lst,Goodhew:2021oqg}, which are more based on locality and unitary of the theory but not directly related to the cosmological collider signals (the branch cuts of the correlators). Exploring the hidden connections among all these cutting rules would be interesting.} Therefore, 
all the signals (both local and nonlocal) in the region $k_{12}>k_{34}$ come from the factorized parts $\mathcal J^{p_1p_2}_{\text{F},>}$, while the time-ordered part $\mathcal J^{p_1p_2}_{\text{TO},>}$ will contribute to the background piece.\footnote{We define the background piece to be analytic in $r_{1,2}$ up to the overall factor $r_1^{p_1}r_2^{p_2}$. When $p_{1,2}$ are not integers, the background piece could also generate an oscillatory shape as in \cite{Chen:2022vzh}.}
However, unlike the tree-level cases, the inverse is not true: As we will see in Sec.\ \ref{sec_bgI}, the factorized part \eqref{eq_SPMB} could also give rise to a background piece, which is convergent and thus free from regularization. This is consistent with the calculation via spectral decomposition \cite{Xianyu:2022jwk}, where the spectral integral of the local signal part of the tree seed can also contribute to the background piece of the loop seed integral.

Once we obtained the PMB representation \eqref{eq_JPMB}-\eqref{eq_BGPMB} for the loop seed integral \eqref{eq_loopseed}, we are left with the multi-layered Mellin integral, and we will demonstrate how to finish the calculation in Sec.\ \ref{sec_calculation} in detail.

\section{Computation of Mellin Integral}
\label{sec_calculation}
In this section, we will calculate the factorized part \eqref{eq_SPMB} and the time-ordered part \eqref{eq_BGPMB} of the loop seed integral, respectively.
Readers uninterested in the technical details can directly go to Sec.\ \ref{sec_result} for the final results.
As we shall see, the loop seed integral $\mathcal J^{p_1p_2}(r_1,r_2)$ breaks into three different pieces according to their analytical behaviors of the kinematics ratios in the squeezed limit $r_{1,2}\to 0$. Schematically, we will have:
\bge
\label{eq_split}
\mathcal J^{p_1p_2}(r_1,r_2) \sim \mathcal G_\text{NS}(r_1,r_2)(r_1r_2)^{\pm\ii\omega} + \mathcal G_\text{LS}(r_1,r_2)\Big(\FR{r_1}{r_2}\Big)^{\pm\ii\omega} +
\mathcal G_\text{BG}(r_1,r_2),
\ede
where $\mathcal G_\text{NS}$, $\mathcal G_\text{LS}$ and $\mathcal G_\text{BG}$ are analytic functions around $r_{1,2} = 0$. The three terms above are called the \emph{nonlocal signal}, the \emph{local signal}, and the \emph{background}, respectively. Both nonlocal and local signals are distinct CC signals, generating logarithmic oscillations with respective to different momentum ratios, namely $r_1r_2=k_s^2/(k_{12}k_{34})$ and $r_1/r_2=k_{34}/k_{12}$, thus are nonanalytic and analytic in $k_s$ around the squeezed limit, respectively. Consequently, after the Fourier transform, they manifest as nonlocal and local in the conjugate spatial distance, thereby validating their names. Notice that the local signal is still nonanalytic in $k_{12}/k_{34}$, while the background piece is fully analytic in all kinematic variables in the squeezed limit.

\subsection{Factorized part}
\label{sec_JF}
Plugging the explicit expressions \eqref{eq_Gamma}, \eqref{eq_Lresult} and \eqref{eq_TF} into \eqref{eq_SPMB}, the factorized piece reads:
\begin{align}
\mathcal J_{\text{F},>}^{p_1p_2} =&~ \FR{r_1^{4+p_1}r_2^{4+p_2}}{64\pi^{7/2}} \int_{-\ii\infty}^{+\ii\infty} \Big[\prod_{i=1}^4\FR{\di s_i}{2\pi\ii}\Big]\,
\bigg[\cos \FR{\pi \bar p_{12}}2 -\cos\Big(2\pi s_{13} - \FR{\pi p_{12}}2\Big)\bigg] \Big(\FR{r_1}2\Big)^{-2s_{13}}\Big(\FR{r_2}2\Big)^{-2s_{24}}\n\\
&\times \Gamma\Big[ 4+p_1-2s_{13},4+p_2-2s_{24}\Big] \times 
\Gamma\left[\bgm
s_{1234}-\fr32,\fr32-s_{12},\fr32-s_{34}\\ 3-s_{1234},s_{12},s_{34}
\edm\right]\n\\
& \times \Gamma\Big[ s_1+\FR{\ii\wt\nu_1}2,s_1-\FR{\ii\wt\nu_1}2,s_2+\FR{\ii\wt\nu_1}2,s_2-\FR{\ii\wt\nu_1}2\Big]\n\\
&\times \Gamma\Big[s_3+\FR{\ii\wt\nu_2}2,s_3-\FR{\ii\wt\nu_2}2,s_4+\FR{\ii\wt\nu_2}2,s_4-\FR{\ii\wt\nu_2}2\Big].
\end{align}
We will finish the Mellin integral using the residue theorem. Since for the physical region, $r_{1,2}<1$, we should close the contour from the left and pick up residues at the left poles of all these Mellin variables.
We immediately find there are four sets of \emph{spectrum poles} for each Mellin variable from the factor $\wt\Gamma(s_1,s_2;s_3,s_4)$ \eqref{eq_Gamma}, namely:
\bge
\label{eq_Specpole}
s_1 = -n_1 - \cc_1\FR{\ii\wt\nu_1}2,\qquad s_2 = -n_2 - \cc_2\FR{\ii\wt\nu_1}2,\qquad s_3 = -n_3 - \cc_3\FR{\ii\wt\nu_2}2,\qquad s_4 = -n_4 - \cc_4\FR{\ii\wt\nu_2}2,
\ede
where $n_i$ take values in nonnegative integers: $n_i=0,1,\cdots$, and $\cc_i=\pm$ (i=1,2,3,4). There is an extra set of \emph{loop UV poles} from the factor $\Gamma(s_{1234}-3/2)$ in the loop integral \eqref{eq_Lresult}:
\bge
\label{eq_UVpole}
s_{1234} = \FR32 -m,\qquad m=0,1,\cdots.
\ede
Readers can refer to \cite{Qin:Local} for the general pole structure of the Mellin integrand. Notice that the loop UV poles \eqref{eq_UVpole} mix the four Mellin variables, while there is an independent series of spectrum poles for each Mellin variable. Therefore, we should carefully consider which poles to select. We can classify the selection by whether we take the loop UV poles \eqref{eq_UVpole} or not.
\subsubsection{Nonlocal signal}
\label{eq_nonlocal}
If we do not select the loop UV poles, we must collect all the spectrum poles \eqref{eq_Specpole} for each Mellin variable. However, the factor $\Gamma(s_{12})\Gamma(s_{34})$ in the denominator of the loop integral forces that we must have $\cc_2=\cc_1$ and $\cc_4=\cc_3$, otherwise the residues vanish. Therefore, the sum of (nonvanishing) residues on the spectrum poles \eqref{eq_Specpole} is:
\begin{align}
\label{eq_JI}
\mathcal J_\text{I}^{p_1p_2} =&~  \FR{r_1^{4+p_1}r_2^{4+p_2}}{64\pi^{7/2}} \sum_{\cc_1,\cc_3=\pm}
\bigg\{\cos \FR{\pi \bar p_{12}}2 -\cos\Big[\pi(\cc_1\ii\wt\nu_1+\cc_3\ii\wt\nu_2)+ \FR{\pi p_{12}}2\Big]\bigg\} \Big(\FR{r_1r_2}4\Big)^{\cc_1\ii\wt\nu_1+\cc_3\ii\wt\nu_2}\n\\
&\times \sum_{n_1,\cdots,n_4=0}^\infty \FR{(-1)^{n_{1234}}}{n_1!n_2!n_3!n_4!}\Big(\FR{r_1}2\Big)^{2n_{13}}\Big(\FR{r_2}2\Big)^{2n_{24}}\n\\
&\times \Gamma\Big[ 4+p_1+2n_{13}+\cc_1\ii\wt\nu_1+\cc_3\ii\wt\nu_2,4+p_2+2n_{24}+\cc_1\ii\wt\nu_1+\cc_3\ii\wt\nu_2\Big]\n\\
& \times 
\Gamma\left[\bgm
-n_{1234}-\cc_1\ii\wt\nu_1-\cc_3\ii\wt\nu_2-\fr32,\fr32+n_{12}+\cc_1\ii\wt\nu_1,\fr32+n_{34}+\cc_3\ii\wt\nu_2\\ 3+n_{1234}+\cc_1\ii\wt\nu_1+\cc_3\ii\wt\nu_2,-n_{12}-\cc_1\ii\wt\nu_1,-n_{34}-\cc_3\ii\wt\nu_2
\edm\right]\n\\
& \times \Gamma\Big[ -n_1 -\cc_1\ii\wt\nu_1, -n_2 -\cc_1\ii\wt\nu_1, -n_3 -\cc_3\ii\wt\nu_2, -n_4 -\cc_3\ii\wt\nu_2\Big].
\end{align}
As one can see, the nonanalytic behavior in terms of $(r_1r_2)^{\pm\ii\omega} \propto k_s^{\pm2\ii\omega}$ with $\omega=|\wt\nu_1\pm\wt\nu_2|$ implies that this piece is the \emph{nonlocal signal}:
\bge
\mathcal J_\text{NS}^{p_1p_2}  = \mathcal J_\text{I}^{p_1p_2}.
\ede
We comment that a similar calculation for the equal-mass case was carried out in \cite{Qin:2022lva}, where we must have $\cc_1=\cc_3$ for the signal in this degenerate case.
\subsubsection{Local signal}
\label{sec_local}
On the other hand, if we pick the loop UV poles \eqref{eq_UVpole}, or equivalently, $s_4=3/2-m-s_{123}$, we can integrate over $s_4$ by collecting residues at these poles and obtain:
\begin{align}
\label{eq_JII}
\mathcal J_\text{II}^{p_1p_2}= &~ \FR{r_1^{4+p_1}r_2^{1+p_2}}{8\pi^{7/2}} \int_{-\ii\infty}^{+\ii\infty} \Big[\prod_{i=1}^3\FR{\di s_i}{2\pi\ii}\Big]\,
\bigg[\cos \FR{\pi \bar p_{12}}2 -\cos\Big(2\pi s_{13} - \FR{\pi p_{12}}2\Big)\bigg]\Big(\FR{r_1}{r_2}\Big)^{-2s_{13}}\n\\
&\times
 \sum_{m=0}^\infty \FR{(-1)^m}{m!}
 \Big(\FR{r_2}2\Big)^{2m}
\times\Gamma\Big[ 4+p_1-2s_{13},1+p_2+2m+2s_{13}\Big]
\n\\
& \times \FR{(s_{12})_m(\fr32-m-s_{12})_m}{\Gamma(\fr32+m)} \times \wt\Gamma\Big( s_1,s_2;s_3,\FR32-m-s_{123}\Big).
\end{align}

Now let us try to integrate over $s_2$.
We recognize the relevant integral involving $s_2$ is:
\begin{align}
\label{eq_Fm}
f_m(s_1,s_3) \equiv &\int_{-\ii\infty}^{+\ii\infty} \FR{\di s_2}{2\pi\ii}\,(s_{12})_m\Big(\FR32-m-s_{12}\Big)_m\n\\
&\times \Gamma\Big[
s_2+\FR{\ii\wt\nu_1}2,s_2-\FR{\ii\wt\nu_1}2, \FR32-m-s_{123} +\FR{\ii\wt\nu_2}2,\FR32-m-s_{123} -\FR{\ii\wt\nu_2}2
 \Big].
\end{align}
 This integral can be completed with the help of Barnes' lemma \eqref{eq_Barnes} \cite{bailey1935generalized}, but with some generalization. We put the calculation details in App.\ \ref{app_fm}, and below is the result:
\begin{align}
\label{eq_fm}
f_m(s_1,s_3) =&~  \sum_{t_1,t_3=0}^m (-1)^{t_{13}}\binom m{t_1}\binom m{t_3}\times
\FR{(1- s_1 -\cc_1\fr{\ii\wt\nu_1}2 -t_1)_{t_1}(1- s_3 -\cc_3\fr{\ii\wt\nu_2}2 -t_3)_{t_3}}{\Gamma(3-2s_{13}-t_{13})}
\n\\
&\times \Gamma\Big[ \FR32-m -s_{13} + \FR{\cc_1\ii\wt\nu_1+\cc_3\ii\wt\nu_2}{2},
\FR32+m-s_{13}- \FR{\cc_1\ii\wt\nu_1+\cc_3\ii\wt\nu_2}{2} -t_{13}\Big]\n\\
&\times \Gamma\Big[
 \FR32 -s_{13} + \FR{\cc_1\ii\wt\nu_1-\cc_3\ii\wt\nu_2}{2}-t_3,
\FR32-s_{13}- \FR{\cc_1\ii\wt\nu_1-\cc_3\ii\wt\nu_2}{2} -t_1
\Big].
\end{align}

Once we complete the integral over $s_2$, we encounter the double-layered integral of $s_1$ and $s_3$, as the following:
\begin{align}
\label{eq_JIInew}
\mathcal J_\text{II}^{p_1p_2}= &~ \FR{r_1^{4+p_1}r_2^{1+p_2}}{8\pi^{7/2}} \int_{-\ii\infty}^{+\ii\infty} \Big[\prod_{i=1,3}\FR{\di s_i}{2\pi\ii}\Big]\,
\bigg[\cos \FR{\pi \bar p_{12}}2 -\cos\Big(2\pi s_{13} - \FR{\pi p_{12}}2\Big)\bigg]\Big(\FR{r_1}{r_2}\Big)^{-2s_{13}}\n\\
&\times
 \sum_{m=0}^\infty \FR{(-1)^m}{m!}
 \Big(\FR{r_2}2\Big)^{2m}
\times\Gamma\Big[ 4+p_1-2s_{13},1+p_2+2m+2s_{13}\Big] \times \FR{f_m(s_1,s_3)}{\Gamma(\fr32+m)}
\n\\
& \times \Gamma\Big[ s_1+\FR{\ii\wt\nu_1}2,s_1-\FR{\ii\wt\nu_1}2,s_3+\FR{\ii\wt\nu_2}2,s_3-\FR{\ii\wt\nu_2}2\Big].
\end{align}
Again, since we focus on the kinematic region that $r_1<r_2$, we should take the right poles of $s_1$ and $s_3$. Since $f_m(s_1,s_3)$ does not contain right poles of $s_1$ and $s_3$, this time we also have two choices:
\begin{enumerate}
\item Choose the spectrum poles \eqref{eq_Specpole}, namely:
\bge
\label{eq_pole1}
s_1 = -n_1 - \cc_1 \FR{\ii\wt\nu_1}2,\qquad s_3 = -n_3 - \cc_3 \FR{\ii\wt\nu_1}2.
\ede
\item Choose the poles from the factor $\Gamma(1+p_2+2m+2s_{13})$ in the time integral:
\bge
\label{eq_pole2}
2s_{13} = -1-p_2-2m-n,\quad n = 0,1,\cdots.
\ede
\end{enumerate}
Notice that the time integral \eqref{eq_T} originally only gives rise to the right poles that are manifest in the explicit expressions \eqref{eq_TF} and \eqref{eq_TTO}. However, once we select the loop UV poles \eqref{eq_UVpole} and thus fix the sum $s_{1234}$, the left poles of $s_{24}$ in the time integral turns to the right poles of $s_{13}$ \eqref{eq_pole2}, which we must take into account when we integrate over $s_1$ and $s_3$. 
By looking at the kinematic dependence $(r_1/r_2)^{-2s_{13}}$ in the expression \eqref{eq_JII} or \eqref{eq_JIInew}, we find that the spectrum poles \eqref{eq_pole1} contribute to the \emph{local signal} which oscillates in the logarithmic scale with frequencies $\omega=|\wt\nu_1\pm\wt\nu_2|$ as the nonlocal signal, but analytic in $k_s$; while the time IR poles \eqref{eq_pole2} contribute to the \emph{background} piece without any nonanalyticity. Let us first focus on the local signal below, and leave the calculation of the background piece from $\mathcal J_\text{II}^{p_1p_2}$ later in Sec.\ \ref{sec_bgI}.

For the local signal piece, summing up the residues of $\mathcal J_\text{II}^{p_1p_2}$ \eqref{eq_JIInew} at the spectrum poles \eqref{eq_pole1}, we obtain:
\begin{align}
\label{eq_JLS>}
\mathcal J_{\text{LS},>}^{p_1p_2}= &~ \FR{r_1^{4+p_1}r_2^{1+p_2}}{8\pi^{7/2}} \sum_{\cc_1,\cc_3=\pm}\bigg\{\cos \FR{\pi \bar p_{12}}2 -\cos\Big[\pi(\cc_1\ii\wt\nu_1+\cc_2\ii\wt\nu_2)+\FR{\pi p_{12}}2\Big]\bigg\} \Big(\FR{r_1}{r_2}\Big)^{\cc_1\ii\wt\nu_1+\cc_3\ii\wt\nu_3}\n\\
&\times 
\sum_{m,n_1,n_3=0}^\infty \FR{(-1)^{m+n_{13}}}{m!n_1!n_3!}
\Big(\FR{r_2}2\Big)^{2m}\Big(\FR{r_1}{r_2}\Big)^{2n_{13}}\times \Gamma\left[\bgm -n_1-\cc_1\ii\wt\nu_1,-n_3-\cc_3\ii\wt\nu_2\\ \fr32+m \edm\right]\n\\
&\times
\Gamma\Big[4+p_1+2n_{13}+\cc_1\ii\wt\nu_1+\cc_3\ii\wt\nu_2,1+p_2+2m-2n_{13}-\cc_1\ii\wt\nu_1-\cc_3\ii\wt\nu_2\Big]\n\\
&\times \sum_{t_1,t_3=0}^m (-1)^{t_{13}}\binom m{t_1}\binom m{t_3}\times
\FR{(1+n_1-t_1)_{t_1}(1+n_2-t_3)_{t_3}}{\Gamma(3+2n_{13}-t_{13}+\cc_1\ii\wt\nu_1+\cc_3\ii\wt\nu_2)}
\n\\
&\times \Gamma\Big[ \FR32-m +n_{13} + \cc_1\ii\wt\nu_1+\cc_3\ii\wt\nu_2,
\FR32+m+n_{13}-t_{13}\Big]\n\\
&\times \Gamma\Big[
 \FR32+n_{13} +\cc_1\ii\wt\nu_1-t_3,
\FR32+n_{13}+\cc_3\ii\wt\nu_2-t_1
\Big],
\end{align}
where we have plugged in the expression of $f_m(s_1,s_3)$ given in \eqref{eq_fm}.
Similar to the nonlocal signal, for the equal-mass case we are forced to have $\cc_1=\cc_3$ to have the signal.

\subsubsection{Background from factorized part}
\label{sec_bgI}
In Sec.\ \ref{sec_local} we have only calculated the local signal part  of $\mathcal J^{p_1p_2}_\text{II}$ \eqref{eq_JIInew} that collecting spectrum poles \eqref{eq_pole1} for $s_1$ and $s_3$. The rest contribution comes from residues \eqref{eq_pole2} from the time integral, and we will see that this remaining contribution is a part of the background piece of the loop seed integral \eqref{eq_loopseed}.

Let us first make the following change of measurement:
\bge
\label{eq_change}
\int_{-\ii\infty}^{+\ii\infty}\FR{\di s_1}{2\pi\ii}\FR{\di s_3}{2\pi\ii} = \int_{-\ii\infty}^{+\ii\infty}\FR{\di S}{2\pi\ii}\FR{\di s_1}{2\pi\ii},\qquad S\equiv s_{13}.
\ede
Then the integrand of \eqref{eq_JIInew} can be rewritten as a function of $S$ and $s_1$:
\begin{align}
\label{eq_JIInew2}
\mathcal J_\text{II}^{p_1p_2}= &~ \FR{r_1^{4+p_1}r_2^{1+p_2}}{8\pi^{7/2}} \int_{-\ii\infty}^{+\ii\infty} \FR{\di S}{2\pi\ii}\FR{\di s_1}{2\pi\ii}\,
\bigg[\cos \FR{\pi \bar p_{12}}2 -\cos\Big(2\pi S - \FR{\pi p_{12}}2\Big)\bigg]\Big(\FR{r_1}{r_2}\Big)^{-2S}\n\\
&\times
 \sum_{m=0}^\infty \FR{(-1)^m}{m!}
 \Big(\FR{r_2}2\Big)^{2m}
\times\Gamma\Big[ 4+p_1-2S,1+p_2+2m+2S\Big] \times \FR{f_m(s_1,S-s_1)}{\Gamma(\fr32+m)}
\n\\
& \times \Gamma\Big[ s_1+\FR{\ii\wt\nu_1}2,s_1-\FR{\ii\wt\nu_1}2,S-s_1+\FR{\ii\wt\nu_2}2,S-s_1-\FR{\ii\wt\nu_2}2\Big].
\end{align}
Now we need to collect residues at poles \eqref{eq_pole2}, namely:
\bge
S = -\FR{1+p_2+2m+n}2,\qquad n=0,1,\cdots,
\ede
and the sum is:
\begin{align}
\mathcal J_\text{III}^{p_1p_2} =&~ \FR{r_1^{5+p_{12}}}{8\pi^{7/2}}
 \sum_{m,n=0}^\infty 
\Big[ \cos\FR{\pi \bar p_{12}}2+ (-1)^n\cos\FR{\pi(p_1+3p_2)}2\Big]
 \FR{(-1)^{m}}{m!} \FR{(-1)^n}{2n!}\n\\
 &\times
 \Big(\FR{r_2}2\Big)^{2m} \Big(\FR{r_1}{r_2}\Big)^{2m+n}\times
\Gamma\left[\bgm 5+p_{12}+2m+n\\\fr32+m\edm \right] \times\int_{-\ii\infty}^{+\ii\infty}\FR{\di s_1}{2\pi\ii}\, f_m(s_1,S-s_1)
\n\\
& \times \Gamma\Big[s_1+\FR{\ii\wt\nu_1}2,s_1-\FR{\ii\wt\nu_1}2,S-s_1+\FR{\ii\wt\nu_2}2, S-s_1-\FR{\ii\wt\nu_2}2\Big]\bigg|_{S=-(1+p_2+2m+n)/2}.
\end{align}
As one can see, the dependence on ratios $r_{1,2}$ is fully analytic, so $\mathcal J_\text{III}^{p_1p_2}$ belongs to the background piece of the loop seed integral. Finally, the remaining integral of $s_1$ can be completed using the Barnes' lemma \eqref{eq_Barnes}, similar to the integral \eqref{eq_Fm}:
\bge
\label{eq_Gm}
g_{m}(S) \equiv
\int_{-\ii\infty}^{+\ii\infty}\FR{\di s_1}{2\pi\ii}\, f_m(s_1,S-s_1) \Gamma\Big[ s_1+\FR{\ii\wt\nu_1}2,s_1-\FR{\ii\wt\nu_1}2,S-s_1+\FR{\ii\wt\nu_2}2, S-s_1-\FR{\ii\wt\nu_2}2\Big],
\ede
where we put the calculation of $g_m(S)$ in App.\ \ref{app_gm} and the final expression of $g_m(S)$ is given in \eqref{eq_gm}:
\begin{align}
g_m(S) =&~ 
\Gamma\Big[ \FR32-m -S + \FR{\ii\wt\nu_1+\ii\wt\nu_2}{2},S-\FR{\ii\wt\nu_1+\ii\wt\nu_2}2\Big]
\sum_{t_1,t_3=0}^m \FR{\binom m{t_1}\binom m{t_3}}{\Gamma[2S+t_{13},3-2S-t_{13}]}\n\\
&\times \Gamma\Big[
 \FR32 -S + \FR{\ii\wt\nu_1-\ii\wt\nu_2}{2}-t_3,
\FR32-S- \FR{\ii\wt\nu_1-\ii\wt\nu_2}{2} -t_1,\FR32+m-S- \FR{\ii\wt\nu_1+\ii\wt\nu_2}{2} -t_{13}
\Big]\n\\
&\times
 \Gamma\left[\bgm 
S+\FR{\ii\wt\nu_1+\ii\wt\nu_2}2+t_{13},S+\FR{\ii\wt\nu_1-\ii\wt\nu_2}2+t_1,S-\FR{\ii\wt\nu_1-\ii\wt\nu_2}2+t_3
\edm\right].
\end{align}

Therefore, we have:
\begin{align}
\label{eq_JIII}
\mathcal J_\text{III}^{p_1p_2} =&~ \FR{r_1^{5+p_{12}}}{8\pi^{7/2}}
\sum_{m,n=0}^\infty \Big[\cos\FR{\pi \bar p_{12}}2 + (-1)^n\cos\FR{\pi(p_1+3p_2)}2\Big] 
 \FR{(-1)^{m}}{m!} \FR{(-1)^n}{2n!}\n\\
 &\times
 \Big(\FR{r_2}2\Big)^{2m} \Big(\FR{r_1}{r_2}\Big)^{2m+n}
 \times
\Gamma\left[\bgm 5+p_{12}+2m+n\\\fr32+m\edm \right] \times g_m\Big(-\FR{1+p_2+2m+n}2\Big).
\end{align}
This finishes our calculation of the factorized part $\mathcal J_{\text{F},>}^{p_1p_2}$ of the loop seed integral \eqref{eq_loopseed}:
\begin{align}
\mathcal J_{\text{F},>}^{p_1p_2} =&~\mathcal J_{\text{I}}^{p_1p_2}+\mathcal J_\text{II}^{p_1p_2}\n\\
=&~ \mathcal J_{\text{NS}}^{p_1p_2}
+( \mathcal J_{\text{L},>}^{p_1p_2}+ \mathcal J_{\text{III}}^{p_1p_2}),
\end{align}
with $\mathcal J_{\text{NS}}^{p_1p_2}=\mathcal J_{\text{I}}^{p_1p_2}$, $\mathcal J_{\text{L},>}^{p_1p_2}$ and $\mathcal J_{\text{III}}^{p_1p_2}$ given in \eqref{eq_JI}, \eqref{eq_JLS>} and \eqref{eq_JIII}, respectively.

At the end of this subsection, we emphasize that $\mathcal J_\text{III}^{p_1p_2}$ \eqref{eq_JIII}, as a part of the background piece of the loop seed integral, is a convergent series (in the region $0<r_1<r_2<1$) without UV divergence. Furthermore, 
notice that $g_m(S)$ has a series of zeros \eqref{eq_zeros}, we immediately know that if $p_2$ is an integer and $p_2\geq -2$ which is the most cases we encounter as explained in Sec.\ \ref{sec_loopseed}, the only case that $\mathcal J_\text{III}^{p_1p_2}$ does not vanish is $p_2=-2$, where only terms with $n=0$ in \eqref{eq_JIII} do not vanish in the summation, and the  expression for this piece can be simplified to:
\begin{align}
\label{eq_JIII-2}
\mathcal J_\text{III}^{p_1,-2} =&- \FR{r_1^{3+p_1}}{8\pi^{7/2}}
\cos\Big(\FR{\pi p_1}2\Big)
 \sum_{m}^\infty \FR{(-1)^{m}}{m!}
 \Big(\FR{r_1}2\Big)^{2m}
\Gamma\left[\bgm 3+p_1+2m\\\fr32+m\edm \right] \times g_m\Big(\FR12-m\Big)\n\\
=&- \FR{r_1^{3+p_1}}{8\pi^{7/2}}
\cos\Big(\FR{\pi p_1}2\Big)
 \sum_{m}^\infty \FR{(-1)^{m}}{m!}
 \Big(\FR{r_1}2\Big)^{2m}
\Gamma\left[\bgm 3+p_1+2m\\\fr32+m\edm \right] \times \FR{(-1)^m 2\pi^4(\wt\nu_1^2-\wt\nu_2^2)}{\cosh2\pi\wt\nu_1-\cosh2\pi\wt\nu_2}\n\\
=&- \FR{1}{4}\cos\Big(\FR{\pi p_1}2\Big)\Gamma(2+p_1)\bigg[\Big(\FR{r_1}{1-r_1}\Big)^{2+p_1}-\Big(\FR{r_1}{1+r_1}\Big)^{2+p_1}\bigg]
\FR{\wt\nu_1^2-\wt\nu_2^2}{\cosh2\pi\wt\nu_1-\cosh2\pi\wt\nu_2},
\end{align}
and in particular,
\bge
\mathcal J_\text{III}^{-2,-2} = \FR{1}{4}\FR{\wt\nu_1^2-\wt\nu_2^2}{\cosh2\pi\wt\nu_1-\cosh2\pi\wt\nu_2}\log\FR{1+r_1}{1-r_1}.
\ede

\subsection{Time-ordered part}
\label{sec_JTO}
Now we turn to the time-ordered part $\mathcal J^{p_1p_2}_{\text{TO},>}$ of the loop seed integral. Plugging \eqref{eq_Gamma}, \eqref{eq_Lresult} and \eqref{eq_TTO} into \eqref{eq_BGPMB}, we obtain:
\begin{align}
\mathcal J_{\text{TO},>}^{p_1p_2} =&~ \FR{r_1^{8+p_{12}}}{32\pi^{7/2}} \int_{-\ii\infty}^{+\ii\infty} \Big[\prod_{i=1}^4\FR{\di s_i}{2\pi\ii}\Big]\,
\sin[\pi(s_{13}-s_{24})]\sin\Big( \FR{\pi p_{12}}2 - \pi s_{1234}\Big)\Big(\FR{r_1}2\Big)^{-2s_{1234}}\n\\
&\times {}_2\mathcal F_1\left[\bgm4+p_2-2s_{24},8+p_{12}-2s_{1234}\\5+p_2-2s_{24}\edm\middle| -\FR{r_1}{r_2}\right] \times 
\Gamma\left[\bgm
s_{1234}-\fr32,\fr32-s_{12},\fr32-s_{34}\\ 3-s_{1234},s_{12},s_{34}
\edm\right]\n\\
& \times \Gamma\Big[ s_1+\FR{\ii\wt\nu_1}2,s_1-\FR{\ii\wt\nu_1}2,s_2+\FR{\ii\wt\nu_1}2,s_2-\FR{\ii\wt\nu_1}2\Big]\n\\
&\times \Gamma\Big[s_3+\FR{\ii\wt\nu_2}2,s_3-\FR{\ii\wt\nu_2}2,s_4+\FR{\ii\wt\nu_2}2,s_4-\FR{\ii\wt\nu_2}2\Big].
\end{align}
Since $r_1<1$, we should take left poles of $s_{1234}$. This time if we take the four spectrum poles \eqref{eq_Specpole} for the Mellin variables, the residues will vanish due to either the factor $\sin[\pi(s_{13}-s_{24})]$ or $1/\Gamma[s_{12},s_{34}]$ in the integrand. Therefore, we must select the loop UV poles \eqref{eq_UVpole}.

The calculation of this piece is very similar to $\mathcal J_\text{III}^{p_1p_2}$:
We first integrate over $s_4$ by summing the residues at the loop UV poles $s_4=3/2-m-s_{123}$, and integrate over $s_2$ using the Barnes' lemma which will give the function $f_m(s_1,s_3)$ \eqref{eq_Fm}. Then we change variables as in \eqref{eq_change} and complete the integral of $s_1$ which will give the function $g_m(S)$ \eqref{eq_Gm}. Finally, we encounter a single-layer Mellin integral over $S\equiv s_{13}$ after some simplifications:
\begin{align}
\label{eq_JTO}
\mathcal J_{\text{TO},>}^{p_1p_2} =&~ \FR{r_1^{5+p_{12}}}{4\pi^{7/2}}\cos\Big(\FR{\pi p_{12}}2\Big) \sum_{m=0}^\infty \FR{(-1)^m}{m!}
 \Big(\FR{r_1}2\Big)^{2m}  \int_{-\ii\infty}^{+\ii\infty} \FR{\di S}{2\pi\ii}\,
 \FR{\cos(2\pi S)}{\Gamma(\fr32+m)}\n\\
&
\times  {}_2\mathcal F_1\left[\bgm
1+p_2+2m+2S,5+p_{12}+2m\\
2+p_2+2m+2S
\edm\middle| -\FR{r_1}{r_2}\right] \times g_m(S).
\end{align}
For later convenience we define the integrand to be:
\bge
\label{eq_wtg}
\wt g_m\Big(S;\FR{r_1}{r_2}\Big) \equiv \FR{\cos(2\pi S)}{\Gamma(\fr32+m)}
\times  {}_2\mathcal F_1\left[\bgm
1+p_2+2m+2S,5+p_{12}+2m\\
2+p_2+2m+2S
\edm\middle| -\FR{r_1}{r_2}\right] \times g_m(S).
\ede

Since \eqref{eq_JTO} is analytic in the two ratios around the limit $0<r_1\ll r_2\ll1$, we recognize this piece also contributes to the \emph{background} piece of the loop seed integral. Therefore, the background piece of the full loop seed integral consists of both this time-ordered part and $\mathcal J_\text{III}^{p_1p_2}$ from the factorized part:
\bge
\mathcal J_{\text{BG},>}^{p_1p_2} = \mathcal J_\text{III}^{p_1p_2}
+\mathcal J_{\text{TO},>}^{p_1p_2}.
\ede

\subsubsection{UV divergence}
The integral in \eqref{eq_JTO}  is divergent.
Let us consider the asymptotic behavior of the integrand as $S\to\pm\ii\infty$. Using the asymptotic behaviors of $\Gamma$ functions, hypergeometric functions and $g_m(S)$, which are given in \eqref{eq_GammaAsym}, \eqref{eq_HyperAsym} and \eqref{eq_Slimit}, respectively, we obtain:
\begin{align}
\label{eq_limitIntegrand}
\lim_{S\to\pm\ii\infty} \wt g_m\Big(S;\FR{r_1}{r_2}\Big)=&~ \FR{e^{2\pi|S|}}{2\Gamma(\fr32+m)}\times \FR{\Gamma(5+p_{12}+2m)}{2S}\Big(\FR{r_2}{r_1+r_2}\Big)^{5+p_{12}+m}\times \lim_{S\to\pm\ii\infty}g_m(S)\n\\
=&~ \begin{cases}
\FR{\ii\pi^{5/2}}{S} \Big(\FR{r_2}{r_1+r_2}\Big)^{5+p_{12}} \Gamma(5+p_{12}),& m=0,\\\mathcal O\Big(\FR{1}{S^2}\Big),&m=1,2,\cdots.
\end{cases}
\end{align}
Since the integral of $1/S$ is divergent in infinity,\footnote{There is a caveat: One may suspect the principal value of the integral $\int_{-\ii\infty}^{\ii\infty} \di S/S = \ii\pi$ is finite, so the integral is free of UV divergence. However, we cannot determine how $S$ approaches $+\ii\infty$ and $-\ii\infty$ in priority. Therefore, we should treat the divergence of the integral at $\pm\ii\infty$ independently and thus they cannot be canceled. Another interesting observation is that the divergence does not arise from the collision of poles in the integrand as one might expect, since this is typically the case in flat spacetime calculations. Further effort is needed to fully understand the underlying reason.} while the integral of $\mathcal O(1/S^2)$ converges, only the term with $m=0$ contributes to the divergence of $\mathcal J^{p_1p_2}_{\text{TO},>}$ \eqref{eq_JTO}. Furthermore, we can extract its divergent part through \eqref{eq_limitIntegrand}:
\begin{align}
\label{eq_JTOdiv}
\mathcal J_{\text{TO},>,\text{div}}^{p_1p_2}=&~ \FR{r_1^{5+p_{12}}}{4\pi^{7/2}}\cos\Big(\FR{\pi p_{12}}2\Big)
\int_{-\ii\infty}^{+\ii\infty} \FR{\di S}{2\pi\ii}\,\FR{\ii\pi^{5/2}}{S} \Big(\FR{r_2}{r_1+r_2}\Big)^{5+p_{12}} \Gamma(5+p_{12})\n\\
 =&~ \FR{1}{8\pi^2}\Big(\FR{r_1r_2}{r_1+r_2}\Big)^{5+p_{12}}\cos\Big(\FR{\pi p_{12}}2\Big)\Gamma(5+p_{12})\int_{-\ii\infty}^{\ii\infty} \FR{\di S}{S}.
\end{align}

Since $\mathcal J_{\text{TO},>}^{p_1p_2}$ is the only piece of the loop seed integral $\mathcal J^{p_1p_2}$ that diverges, \eqref{eq_JTOdiv} is also the divergent part of the full loop seed integral:
\bge
\label{eq_Jdiv}
\mathcal J_{\text{div}}^{p_1p_2}(r_1,r_2) = \mathcal J_{\text{TO},>,\text{div}}^{p_1p_2}(r_1,r_2)
\ede

\subsubsection{Renormalization}
Notice that the divergent part of the loop seed integral \eqref{eq_Jdiv}, solely coming from the time-ordered integral \eqref{eq_JTOdiv}, 
depends on the kinematic variables via the combination $r_1r_2/(r_1+r_2) = k_s/k_{1234}$. This particular kinematic dependence (shape) is the same as a four-point correlator generated from a contact graph.
Therefore, the UV divergence of the 1-loop correlator is completely local and can be subtracted by a local counterterm.

To be more specific, let us consider the four-point correlator \eqref{eq_massless4ptgeneral} generated from the bubble diagram Fig.\ \ref{fig_bubble} with the interaction \eqref{eq_masslessL}, then its divergent part can be easily read from \eqref{eq_massless4ptgeneral}:
\bge
\label{eq_divbubble}
\la \varphi_{\mb k_1}\varphi_{\mb k_2}\varphi_{\mb k_3}\varphi_{\mb k_4} \ra'_{s,\text{div}}
\supset \FR{1}{16k_1k_2k_3k_4k_s^{5+p_{12}}} \mathcal J_{\text{div}}^{p_1p_2}(r_1,r_2).
\ede
On the other hand, we can compute the four-point correlator generated from the contact graph with quartic interaction:
\bge
\label{eq_Lcount}
\Delta \ld_\text{c} = -\FR{\lam_*}{4!}(-\tau)^{p_{12}}\varphi'^4.
\ede
The result reads:
\begin{align}
\label{eq_counter}
\la \varphi_{\mb k_1}\varphi_{\mb k_2}\varphi_{\mb k_3}\varphi_{\mb k_4} \ra'_{\text{c}} = & - \lam_* \sum_{\aa=\pm}\ii\aa\int_{-\infty}^0\di\tau\,(-\tau)^{p_{12}}G'_\aa(k_1;\tau)G'_\aa(k_2;\tau)G'_\aa(k_3;\tau)G'_\aa(k_4;\tau)\n\\
=&-\FR{\lam_*}{8k_1k_2k_3k_4k_s^{5+p_{12}}}\cos\Big(\FR{\pi p_{12}}2\Big)\Gamma(5+p_{12})\Big(\FR{r_1r_2}{r_1+r_2}\Big)^{5+p_{12}}.
\end{align}
By comparing \eqref{eq_counter} with \eqref{eq_divbubble}, we find the UV divergence of the bubble diagram can be completely subtracted by the counterterm \eqref{eq_Lcount}, with the infinite coefficient $\lam_*$ determined by:
\bge
\lam_* = \FR{1}{16\pi^2} \int_{-\ii\infty}^{\ii\infty} \FR{\di S}{S}.
\ede
We also mention that the counterterm \eqref{eq_Lcount} has the same form as in flat spacetime, since the UV divergence comes from the hard loop momentum region where the spacetime curvature can be neglected.

However, we also have the freedom to vary the coefficient in the counterterm by adding any finite number:
\bge
\label{eq_Lcounter}
\Delta \ld_\text{c} = -\FR{\lam_*+c}{4!}(-\tau)^{p_{12}}\varphi'^4,
\ede
since this will not affect the divergent part of the contact graph. Therefore, the divergence from the bubble from interaction \eqref{eq_masslessL} and the divergence from the contact graph from the counterterm \eqref{eq_Lcounter} will still cancel, and the remaining finite result is the \emph{renormalized loop seed integral}, up to the trivial prefactor in \eqref{eq_massless4ptgeneral}.Since the divergence of the loop seed integral only comes from $\mathcal J^{p_1p_2}_{\text{TO},>}$, it is equivalent to have a renormalized $\mathcal J^{p_1p_2}_{\text{TO},>,\text{re}}$, which is:
\begin{align}
\mathcal J^{p_1p_2}_{\text{TO},>,\text{re}} =&~\mathcal J^{p_1p_2}_{\text{TO},>} - (\lam_*+c)\times 2\cos\Big(\FR{\pi p_{12}}2\Big)\Gamma(5+p_{12})\Big(\FR{r_1r_2}{r_1+r_2}\Big)^{5+p_{12}}\n\\
=&~
\FR{r_1^{5+p_{12}}}{4\pi^{7/2}}\cos\Big(\FR{\pi p_{12}}2\Big)\int_{-\ii\infty}^{+\ii\infty} \FR{\di S}{2\pi\ii}\,\bigg[\wt g_0\Big(S,\FR{r_1}{r_2}\Big) - \FR{\ii\pi^{5/2} }S\Big(\FR{r_2}{r_1+r_2}\Big)^{5+p_{12}} \Gamma(5+p_{12})\bigg]
\n\\
&+\FR{r_1^{5+p_{12}}}{4\pi^{7/2}}\cos\Big(\FR{\pi p_{12}}2\Big) \sum_{m=1}^\infty \FR{(-1)^m}{m!}
 \Big(\FR{r_1}2\Big)^{2m}  \int_{-\ii\infty}^{+\ii\infty} \FR{\di S}{2\pi\ii}\,\wt g_m\Big(S,\FR{r_1}{r_2}\Big)\n\\
 &+ C\Big(\FR{r_1r_2}{r_1+r_2}\Big)^{5+p_{12}},
\end{align}
where $C$ is an arbitrary number (coming from $c$ in the counterterm \eqref{eq_Lcounter}) that should be determined by renormalization. In principle, one can continue to evaluate the final fold of the Mellin integral of $S$ by collecting residues at either the left or right poles of the integrand, both of which yield the same convergent result since the divergent part has been canceled by the counterterm. However, we will not pursue this here. This integral can also be efficiently computed numerically.

\subsection{Summary and discussions}
\label{sec_result}
At this point, we have finished the computation of loop seed integral $\mathcal J^{p_1p_2}(r_1,r_2)$ \eqref{eq_loopseed}, and we present the full result below. In particular, we focus on the expression in the region $0<r_1<r_2<1$. The loop seed integral can be split into three pieces according to the analytical behavior of $r_{1,2}$:
\bge
\mathcal J^{p_1p_2}(r_1,r_2) = \mathcal J^{p_1p_2}_\text{NS}(r_1,r_2) +\mathcal J^{p_1p_2}_{\text{L},>}(r_1,r_2) +\mathcal J^{p_1p_2}_{\text{BG},>}(r_1,r_2).
\ede
Both the nonlocal and local signals come from the factorized part of the loop seed integral, given by:
\begin{align}
\label{eq_NSFinal}
\mathcal J_\text{NS}^{p_1p_2}(r_1,r_2) =&~  \FR{r_1^{4+p_1}r_2^{4+p_2}}{64\pi^{7/2}} \sum_{\cc_1,\cc_3=\pm}
\bigg\{\cos \FR{\pi \bar p_{12}}2 -\cos\Big[\pi(\cc_1\ii\wt\nu_1+\cc_3\ii\wt\nu_2)+ \FR{\pi p_{12}}2\Big]\bigg\} \Big(\FR{r_1r_2}4\Big)^{\cc_1\ii\wt\nu_1+\cc_3\ii\wt\nu_2}\n\\
&\times \sum_{n_1,\cdots,n_4=0}^\infty \FR{(-1)^{n_{1234}}}{n_1!n_2!n_3!n_4!}\Big(\FR{r_1}2\Big)^{2n_{13}}\Big(\FR{r_2}2\Big)^{2n_{24}}\n\\
&\times \Gamma\Big[ 4+p_1+2n_{13}+\cc_1\ii\wt\nu_1+\cc_3\ii\wt\nu_2,4+p_2+2n_{24}+\cc_1\ii\wt\nu_1+\cc_3\ii\wt\nu_2\Big]\n\\
& \times 
\Gamma\left[\bgm
-n_{1234}-\cc_1\ii\wt\nu_1-\cc_3\ii\wt\nu_2-\fr32,\fr32+n_{12}+\cc_1\ii\wt\nu_1,\fr32+n_{34}+\cc_3\ii\wt\nu_2\\ 3+n_{1234}+\cc_1\ii\wt\nu_1+\cc_3\ii\wt\nu_2,-n_{12}-\cc_1\ii\wt\nu_1,-n_{34}-\cc_3\ii\wt\nu_2
\edm\right]\n\\
& \times \Gamma\Big[ -n_1 -\cc_1\ii\wt\nu_1, -n_2 -\cc_1\ii\wt\nu_1, -n_3 -\cc_3\ii\wt\nu_2, -n_4 -\cc_3\ii\wt\nu_2\Big],
\end{align}
and
\begin{align}
\label{eq_LSFinal}
\mathcal J_{\text{LS},>}^{p_1p_2}(r_1,r_2)= &~ \FR{r_1^{4+p_1}r_2^{1+p_2}}{8\pi^{7/2}} \sum_{\cc_1,\cc_3=\pm}\bigg\{\cos \FR{\pi \bar p_{12}}2 -\cos\Big[\pi(\cc_1\ii\wt\nu_1+\cc_2\ii\wt\nu_2)+\FR{\pi p_{12}}2\Big]\bigg\} \Big(\FR{r_1}{r_2}\Big)^{\cc_1\ii\wt\nu_1+\cc_3\ii\wt\nu_3}\n\\
&\times 
\sum_{m,n_1,n_3=0}^\infty \FR{(-1)^{m+n_{13}}}{m!n_1!n_3!}
\Big(\FR{r_2}2\Big)^{2m}\Big(\FR{r_1}{r_2}\Big)^{2n_{13}}\times \Gamma\left[\bgm -n_1-\cc_1\ii\wt\nu_1,-n_3-\cc_3\ii\wt\nu_2\\ \fr32+m \edm\right]\n\\
&\times
\Gamma\Big[4+p_1+2n_{13}+\cc_1\ii\wt\nu_1+\cc_3\ii\wt\nu_2,1+p_2+2m-2n_{13}-\cc_1\ii\wt\nu_1-\cc_3\ii\wt\nu_2\Big]\n\\
&\times \sum_{t_1,t_3=0}^m (-1)^{t_{13}}\binom m{t_1}\binom m{t_3}\times
\FR{(1+n_1-t_1)_{t_1}(1+n_2-t_3)_{t_3}}{\Gamma(3+2n_{13}-t_{13}+\cc_1\ii\wt\nu_1+\cc_3\ii\wt\nu_2)}
\n\\
&\times \Gamma\Big[ \FR32-m +n_{13} + \cc_1\ii\wt\nu_1+\cc_3\ii\wt\nu_2,
\FR32+m+n_{13}-t_{13}\Big]\n\\
&\times \Gamma\Big[
 \FR32+n_{13} +\cc_1\ii\wt\nu_1-t_3,
\FR32+n_{13}+\cc_3\ii\wt\nu_2-t_1
\Big],
\end{align}
respectively.

The background comes from both the factorized part ($\mathcal J_\text{III}^{p_1p_2}$) and the time-ordered part ($\mathcal J_{\text{TO},>}^{p_1p_2}$). The former is free from UV divergence, whereas the latter is not; this divergence can be canceled using the contact counterterm \eqref{eq_Lcounter} same as in flat spacetime. The renormalized background is:
\begin{align}
\label{eq_BGFinal}
\mathcal J_{\text{BG},>,\text{re}}^{p_1p_2}(r_1,r_2) =&~ \FR{r_1^{5+p_{12}}}{8\pi^{7/2}}
\sum_{m,n=0}^\infty \Big[ \cos\FR{\pi \bar p_{12}}2 + (-1)^n\cos\FR{\pi(p_1+3p_2)}2\Big] 
 \FR{(-1)^{m}}{m!} \FR{(-1)^n}{2n!}\n\\
 &\times
 \Big(\FR{r_2}2\Big)^{2m} \Big(\FR{r_1}{r_2}\Big)^{2m+n}
 \times
\Gamma\left[\bgm 5+p_{12}+2m+n\\\fr32+m\edm \right] \times g_m\Big(-\FR{1+p_2+2m+n}2\Big)\n\\
&+\FR{r_1^{5+p_{12}}}{4\pi^{7/2}}\cos\Big(\FR{\pi p_{12}}2\Big)\int_{-\ii\infty}^{+\ii\infty} \FR{\di S}{2\pi\ii}\,\bigg[\wt g_0\Big(S;\FR{r_1}{r_2}\Big) - \FR{\ii\pi^{5/2} }S\Big(\FR{r_2}{r_1+r_2}\Big)^{5+p_{12}} \Gamma(5+p_{12})\bigg]
\n\\
&+\FR{r_1^{5+p_{12}}}{4\pi^{7/2}}\cos\Big(\FR{\pi p_{12}}2\Big) \sum_{m=1}^\infty \FR{(-1)^m}{m!}
 \Big(\FR{r_1}2\Big)^{2m}  \int_{-\ii\infty}^{+\ii\infty} \FR{\di S}{2\pi\ii}\,\wt g_m\Big(S;\FR{r_1}{r_2}\Big)\n\\
 &+ C\Big(\FR{r_1r_2}{r_1+r_2}\Big)^{5+p_{12}},
\end{align}
where $g_m$ and $\wt g_m$ are given in \eqref{eq_gm} and \eqref{eq_wtg}, respectively, and $C$ is an aribitrary number determined by renormalization.

\paragraph{Total and partial energy poles}

Though we have specified our kinematic region as $0<r_1<r_2<1$, the entire calculation remains valid after analytical continuation to the region $0<|r_1|<|r_2|<1$. Therefore, we can push our expression toward the boundary from within this region to examine the divergence of each term in the loop seed integral.

For instance, we can fix $0<r_1<1$ in the physical region and push $r_2\to-1$. In this case, we find the signals, namely \eqref{eq_NSFinal} and \eqref{eq_LSFinal} are divergent, since they are series expansions in terms of powers of $r_2$. This is the \emph{partial energy pole} at $r_2=-1$. On the other hand, the background piece does not diverge in this limit, meaning the partial energy pole appears only for the factorized signal parts. We also encounter another partial energy pole when we fix  $0<r_2<1$ and let $r_1\to-1$, since the full result of the loop seed integral must remain invariant under the interchange $(p_1,r_1)\leftrightarrow(p_2,r_2)$. However, we cannot reach this pole directly since the expansions in \eqref{eq_NSFinal}, \eqref{eq_LSFinal} and \eqref{eq_BGFinal} are not valid in this limit and should be analytically continued.

We can also fix $0<r_2<1$ in the physical region and push $r_1\to -r_2$. Now the signals \eqref{eq_NSFinal} and \eqref{eq_LSFinal} remain finite, but the background piece \eqref{eq_BGFinal}, especially the time-ordered part \eqref{eq_JTO} diverges since the argument of the hypergeometric function in the integrand becomes $1$. This is the \emph{total energy pole} at $r_1+r_2=0$.

Interestingly, there may also be a background contribution $\mathcal J_\text{III}^{p_1p_2}$ \eqref{eq_JIII} from the factorized part. As mentioned in Sec.\ \ref{sec_bgI}, this contribution arises only when $p_2=-2$ in most cases we are considering, where there are neither total energy poles nor partial energy poles in its expression \eqref{eq_JIII-2}.
Thus, the partial energy poles arise from the signals of the factorized part, while the total energy pole originates from the time-ordered part, after we decompose the original bubble correlator according to the cutting rules \cite{Tong:2021wai,Qin:Local}. It will be interesting to further explore the analytical structure of cosmological correlators at the loop level, and we hope to address this in future work.

\paragraph{Hierarchical squeezed limit}
From the phenomenological perspective, it is useful to display the leading result in a particular kinematic configuration, namely the hierarchical squeezed limit $k_s\ll k_{34}\ll k_{12}$, or equivalently $r_1\ll r_2\ll1$, where the CC signals could dominate. From the above results, it is easy to write down the leading term of the loop seed integral in this limit, where the nonlocal and local signals are:
\begin{align}
\label{eq_NSlimit}
\lim_{r_1\ll r_2\ll1}\mathcal J_\text{NS}^{p_1p_2}(r_1,r_2) =&~  \FR{r_1^{4+p_1}r_2^{4+p_2}}{64\pi^{7/2}} \sum_{\cc_1,\cc_3=\pm}
\bigg\{\cos \FR{\pi \bar p_{12}}2 -\cos\Big[\pi(\cc_1\ii\wt\nu_1+\cc_3\ii\wt\nu_2)+ \FR{\pi p_{12}}2\Big]\bigg\} \n\\
&\times\Big(\FR{r_1r_2}4\Big)^{\cc_1\ii\wt\nu_1+\cc_3\ii\wt\nu_2} \times \Gamma\left[\bgm
-\fr32-\cc_1\ii\wt\nu_1-\cc_3\ii\wt\nu_2,\fr32+\cc_1\ii\wt\nu_1,\fr32+\cc_3\ii\wt\nu_2\\ 3+\cc_1\ii\wt\nu_1+\cc_3\ii\wt\nu_2
\edm\right],\n\\
&\times\Gamma\Big[ 4+p_1+\cc_1\ii\wt\nu_1+\cc_3\ii\wt\nu_2,4+p_2+\cc_1\ii\wt\nu_1+\cc_3\ii\wt\nu_2,-\cc_1\ii\wt\nu_1, -\cc_3\ii\wt\nu_2\Big]\\
\label{eq_LSlimit}
\lim_{r_1\ll r_2\ll1}\mathcal J_{\text{LS},>}^{p_1p_2}(r_1,r_2)= &~ \FR{r_1^{4+p_1}r_2^{1+p_2}}{8\pi^{7/2}} \sum_{\cc_1,\cc_3=\pm}\bigg\{\cos \FR{\pi \bar p_{12}}2 -\cos\Big[\pi(\cc_1\ii\wt\nu_1+\cc_2\ii\wt\nu_2)+\FR{\pi p_{12}}2\Big]\bigg\}\n\\
&\times
 \Big(\FR{r_1}{r_2}\Big)^{\cc_1\ii\wt\nu_1+\cc_3\ii\wt\nu_3}\times
 \Gamma\left[\bgm  \fr32+ \cc_1\ii\wt\nu_1+\cc_3\ii\wt\nu_2,\fr32+\cc_1\ii\wt\nu_1,
\fr32+\cc_3\ii\wt\nu_2\\
3+\cc_1\ii\wt\nu_1+\cc_3\ii\wt\nu_2
\edm\right]\n\\
&\times\Gamma\Big[4+p_1+\cc_1\ii\wt\nu_1+\cc_3\ii\wt\nu_2,1+p_2-\cc_1\ii\wt\nu_1-\cc_3\ii\wt\nu_2,-\cc_1\ii\wt\nu_1,-\cc_3\ii\wt\nu_2\Big],
\end{align}
respectively,
while the background piece is:
\begin{align}
\lim_{r_1\ll r_2\ll1}\mathcal J_{\text{BG},>,\text{re}}^{p_1p_2}(r_1,r_2) =&~ \FR{r_1^{5+p_{12}}}{8\pi^4}
\Big[ \cos\FR{\pi \bar p_{12}}2 + \cos\FR{\pi(p_1+3p_2)}2 \Big]
\Gamma( 5+p_{12})\times g_0\Big(-\FR{1+p_2}2\Big)\n\\
&+\FR{r_1^{5+p_{12}}}{2\pi^4}\cos\Big(\FR{\pi p_{12}}2\Big)\Gamma( 5+p_{12})\int_{-\ii\infty}^{+\ii\infty} \FR{\di S}{2\pi\ii}\,\Big[\FR{\cos(2\pi S)g_0(S)}{1+p_2+2S}- \FR{\ii\pi^3}{2S}\Big]
\n\\
 &+ Cr_1^{5+p_{12}}.
\end{align}
As one can see, for the massless case that respects dilation symmetry \eqref{eq_massless4pt} where $p_1=p_2=0$, the signal (especially the local signal) dominates in this hierarchical squeezed limit. We can also read out the Boltzmann suppression factor $\exp[-\pi(\wt\nu_1+\wt\nu_2)]$ for the signal in the large mass limit as expected.

\begin{figure}
\centering 
\includegraphics[width=0.6\textwidth]{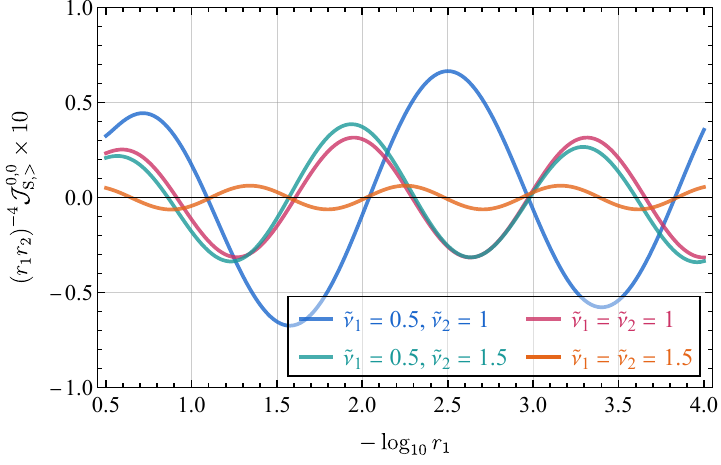} 
\caption{
The signal part of the loop seed integral corresponding to the 1-loop inflaton trispectra \eqref{eq_massless4pt} mediated by a covariant bubble of massive scalars $\si_1$ and $\si_2$ with arbitrary masses $\wt\nu_1$ and $\wt\nu_2$.
In this figure we fix $r_2 = 0.9$ and vary $r_1 \in (10^{-4},10^{-1/2})$. The equal-mass cases $\wt\nu_1=\wt\nu_2=1$ and $\wt\nu_1=\wt\nu_2=1.5$ are consistent with Fig. 4 in \cite{Xianyu:2022jwk}.}
  \label{fig_covariant}
\end{figure}

With the full analytical result, we plot the signal part of the corresponding loop seed integral $\mathcal J^{0,0}_{\text{S},>}\equiv \mathcal J^{0,0}_{\text{NS}}+\mathcal J^{0,0}_{\text{LS},>}$ in Fig.\ \ref{fig_covariant}, with several choices of internal masses $\wt\nu_1$, $\wt\nu_2$ in the bubble. In particular, the equal-mass cases reproduce the results of Fig.\ 4 in \cite{Xianyu:2022jwk}.\footnote{Notice that in Fig.\ 4 of \cite{Xianyu:2022jwk}, the authors only plot the leading term of the signal in the squeezed limit, namely the terms with $n=0$ in equations 86 and 87. We have verified that their full results differ from ours here only by the symmetry factor $1/2$. }
It is interesting when the masses of two internal lines in the bubble are different, i.e., $\wt\nu_1\neq \wt\nu_2$. In this case, we have signals with two possible frequencies, $\omega_\pm = |\wt\nu_1\pm\wt\nu_2|$, corresponding to the choices of signs $\cc_1=\pm \cc_3$. Signals of both frequencies have the same Boltzmann suppression, so their amplitudes are comparable.\footnote{Approximate results for the signals of bispectra have been explored in \cite{Aoki:2020zbj} using the late-time expansion, which have similar phenomenological behaviors. In principle, we can also derive the precise analytical expressions for bispectra by sending $r_2\to1$ as in \cite{Xianyu:2022jwk}. This is not trivial though, because we have to care about the spurious folded pole, and we will leave it to future study.}

\section{Application: Signal from Loop without Full dS Isometries}
\label{sec_break}
One of the significant strengths of the PMB representation is that it does not require full dS isometries for heavy massive fields, unlike the method of spectral decomposition \cite{Xianyu:2022jwk} which heavily relies on them. In this section, we will explicitly present this advantage of the PMB representation by considering a nontrivial example with dS-boost breaking of the massive field, namely the bubble diagram generated by the following interaction:
\bge
\label{eq_boostbreakingL}
\Delta \ld \supset  -\FR14 \varphi'^2 \si'^2.
\ede
This interaction can be generated from the following underlying model: Assume the inflaton derivatively couples to another heavy scalar $\chi$ via $\sqrt{-g}(\pd\varphi)^2\chi^2$, and also assume there is an interaction among the three fields through $\sqrt{-g}\chi(\pd^\mu\varphi)(\pd_\mu\si)$. The latter will generate a two-point mixing between $\chi$ and $\si$ in the form of $a^3 \chi \si'$ when $\varphi$ is taken to the rolling background. Finally, integrating out the heavy scalar $\chi$ will generate the dS-boost-breaking interaction $a^2(\pd \varphi)^2 \si'^2$ that contains \eqref{eq_boostbreakingL}. A more natural dS-boost-breaking model is the loop generated by spinning particles with helical chemical potentials \cite{Chen:2018xck,Liu:2019fag,Wang:2019gbi,Wang:2020ioa,Tong:2022cdz} as we considered in \cite{Qin:2022lva}. However, for our first nontrivial (in the sense that it cannot be solved by any other methods to the best of our knowledge) application of the PMB representation at the loop level, we will focus on the simpler scalar bubble case with the interaction given in \eqref{eq_boostbreakingL}, and we will concentrate on calculating the signals.

It is straightforward to write down the $s$-channel four-point correlator generated by interaction \eqref{eq_boostbreakingL} following the SK formalism:
\begin{align}
\label{eq_boostbreaking4pt}
\la \varphi_{\mb k_1} \varphi_{\mb k_2} \varphi_{\mb k_3} \varphi_{\mb k_4} \ra_s' =&~\FR12\sum_{\aa,\bb=\pm}(-\aa\bb)\int_{-\infty}^0 \di\tau_1\di\tau_2\,G_\aa'(k_1;\tau_1)G_\aa'(k_2;\tau_1)G_\bb'(k_3;\tau_2)G_\bb'(k_4;\tau_2)\n\\
&\times \int\FR{\di^3\mb q}{(2\pi)^3}\,\Big[\pd_{\tau_1}\pd_{\tau_2}D^{(\wt\nu)}_{\aa\bb}(q;\tau_1,\tau_2)\Big]\Big[\pd_{\tau_1}\pd_{\tau_2}D^{(\wt\nu)}_{\aa\bb}(|\mb q+\mb k_s|;\tau_1,\tau_2)\Big].
\end{align}
Here $1/2$ accounts for the symmetry factor, and $\wt\nu$ denotes the mass parameter of $\si$.
For later convenience we use a dimensionless function $\mathcal I(r_1,r_2)$ to express the four-point correlator \eqref{eq_boostbreaking4pt}:
\bge
\la \varphi_{\mb k_1} \varphi_{\mb k_2} \varphi_{\mb k_3} \varphi_{\mb k_4} \ra_s' = \FR{k_s}{16k_1k_2k_3k_4}\mathcal I(r_1,r_2),
\ede
which is the analog of the covariant loop seed integral $\mathcal J^{0,0}(r_1,r_2)$ for the covariant case \eqref{eq_massless4pt}.
Here $r_{1,2}$ are the same ratios of momenta \eqref{eq_ratio}. Without loss of generality, we again assume $0<r_1<r_2<1$. Next, we can easily write down the PMB representation for the derivatives of bulk-to-bulk propagators from \eqref{eq_DPMB}:\footnote{
Strictly speaking, the propagators $\wt D_{\aa\bb}(k_s;\tau_1,\tau_2)$ for $\si'$ are slightly different from directly taking derivatives to propagators for $\si$, namely $\pd_{\tau_1}\pd_{\tau_2}D_{\aa\bb}(k;\tau_1,\tau_2)$: There is an additional term proportional to $\de(\tau_1-\tau_2)$ for the same-sign propagators $\wt D_{\pm\pm}(k_s;\tau_1,\tau_2)$, since taking derivatives does not commute with the $\theta$ functions in the time orderings. However, we focus on the signal parts, while the $\de$ term only gives rise to a contact graph contributing only to the background piece, so we ignore this difference in the following calculations. A similar situation occurs for the longitudinal mode of spin-1 particles, see e.g. \cite{Qin:2022fbv}.}
\begin{align}
\pd_{\tau_1}\pd_{\tau_2}D^{(\wt\nu)}_{\aa\bb}(k;\tau_1,\tau_2) =& \int_{-\ii\infty}^{+\ii\infty} \FR{\di s_1}{2\pi\ii}\FR{\di s_2}{2\pi\ii}\, N_{\aa\bb}(s_1,s_2;\tau_1,\tau_2)(-\tau_1)^{1/2-2s_1}(-\tau_2)^{1/2-2s_2}k^{-2s_{12}}\n\\
&\times \Big(\FR32-2s_1\Big)\Big(\FR32-2s_2\Big)\Gamma\Big[s_1+\FR{\ii\wt\nu}2,s_1-\FR{\ii\wt\nu}2,s_2+\FR{\ii\wt\nu}2,s_2-\FR{\ii\wt\nu}2\Big],
\end{align}
under which the correlator becomes:
\begin{align}
\mathcal I(r_1,r_2)=&~ \FR{k_s}{2} \int_{-\ii\infty}^{+\ii\infty} \Big[\prod_{i=1}^4\FR{\di s_i}{2\pi\ii}\Big]\,\wt{\mathcal T}(k_{12},k_{34})\times\mathcal L(k_s)\times \prod_{i=1}^4\Big(\FR32-2s_i\Big)\times\wt\Gamma(s_1,s_2;s_3,s_4).
\end{align}
Here the time integral $\wt{\mathcal T}(k_{12},k_{34})$ is given in \eqref{eq_Tresult}-\eqref{eq_TTO} with $p_1=p_2=0$:
\begin{align}
\label{eq_T'result}
\wt{\mathcal T}(k_{12},k_{34})  =&~ \wt{\mathcal T}_{\text{F},>}(k_{12},k_{34})+ \wt{\mathcal T}_{\text{TO},>}(k_{12},k_{34}),\\
\label{eq_T'F}
\wt{\mathcal T}_{\text{F},>}(k_{12},k_{34}) =&~ \FR{2^{2s_{1234}}}{8\pi^2}\big[1 -\cos(2\pi s_{13})\big] k_{12}^{-4+2s_{13}} k_{34}^{-4+2s_{24}} \Gamma\Big[ 4-2s_{13},4-2s_{24}\Big],\\
\label{eq_T'TO}
\mathcal T_{\text{TO},>}(k_{12},k_{34})=& - \FR{2^{2s_{1234}}}{4\pi^2} \sin[\pi(s_{13}-s_{24})]\sin(\pi s_{1234}) k_{12}^{-8+2s_{1234}}\n\\&\times {}_2\mathcal F_1\left[\bgm4-2s_{24},8-2s_{1234}\\5-2s_{24}\edm\middle| -\FR{k_{34}}{k_{12}}\right],
\end{align}
the loop integral $\mathcal L(k_s)$ is given in \eqref{eq_Lresult}, and the $\Gamma$ product is given in \eqref{eq_Gamma} with $\wt\nu_1=\wt\nu_2=\wt\nu$:
\bge
\label{eq_Gamma'}
\wt\Gamma(s_1,s_2,s_3,s_4) = \prod_{i=1}^4\Gamma\Big[ s_i+\FR{\ii\wt\nu_1}2,s_i-\FR{\ii\wt\nu_1}2\Big].
\ede
Now, combining \eqref{eq_T'result}-\eqref{eq_T'TO}, \eqref{eq_Lresult} and \eqref{eq_Gamma'}, we can express the correlator $\mathcal I(r_1,r_2)$ in terms of multi-layer Mellin integral:
\begin{align}
\label{eq_IPMB}
\mathcal I(r_1,r_2)  =&~ \mathcal I_{\text{F},>}(r_1,r_2)+ \mathcal I_{\text{TO},>}(r_1,r_2),\\
\label{eq_SI}
\mathcal I_{\text{F},>}(r_1,r_2) =&~ \FR{k_s}{2} \int_{-\ii\infty}^{+\ii\infty} \Big[\prod_{i=1}^4\FR{\di s_i}{2\pi\ii}\Big]\, \wt{\mathcal T}_{\text{F},>}(k_{12},k_{34}) \times \mathcal L(k_s)\n\\
&\times \prod_{i=1}^4\Big(\FR32-2s_i\Big) \wt\Gamma(s_1,s_2,s_3,s_4),\\
\label{eq_BGI}
\mathcal I_{\text{TO},>}(r_1,r_2) =&~ \FR{k_s}{2} \int_{-\ii\infty}^{+\ii\infty} \Big[\prod_{i=1}^4\FR{\di s_i}{2\pi\ii}\Big]\, \wt{\mathcal T}_{\text{TO},>}(k_{12},k_{34}) \times \mathcal L(k_s)\n\\
&\times  \prod_{i=1}^4\Big(\FR32-2s_i\Big)\wt\Gamma(s_1,s_2,s_3,s_4).
\end{align}

Comparing \eqref{eq_IPMB}-\eqref{eq_BGI} with the expressions for the dS-covariant loop seed integral \eqref{eq_JPMB}-\eqref{eq_BGPMB}, we observe that the only difference is the presence of an additional polynomial of Mellin variables in the integrand, which does not affect the pole structure. Consequently, the calculation of the Mellin integral can be performed in a similar manner as the covariant case in Sec.\ \ref{sec_calculation}. In particular, we will present the calculation of the signal part for the remainder of this section, which is contained in the factorized part \eqref{eq_SI}:
\begin{align}
\mathcal I_{\text{F},>}(r_1,r_2)=&~ \FR{r_1^4r_2^4}{128\pi^{7/2}} \int_{-\ii\infty}^{+\ii\infty} \Big[\prod_{i=1}^4\FR{\di s_i}{2\pi\ii}\Big]\,
\big[1 -\cos(2\pi s_{13})\big] \Big(\FR{r_1}2\Big)^{-2s_{13}}\Big(\FR{r_2}2\Big)^{-2s_{24}}\n\\
&\times \Gamma\Big[ 4-2s_{13},4-2s_{24}\Big] \times 
\Gamma\left[\bgm
s_{1234}-\fr32,\fr32-s_{12},\fr32-s_{34}\\ 3-s_{1234},s_{12},s_{34}
\edm\right]\n\\
&\times \prod_{i=1}^4\Big(\FR32-2s_i\Big)\wt\Gamma(s_1,s_2,s_3,s_4).
\end{align}

\paragraph{Nonlocal signal}
The nonlocal signal comes from the contribution of spectrum poles from $\wt\Gamma(s_1,s_2,s_3,s_4)$:
\bge
s_1 = -n_1\mp\ii \cc\wt\nu,\qquad s_2 = -n_2\mp\ii \cc\wt\nu,\qquad s_3 = -n_3\mp\ii \cc\wt\nu,\qquad s_4 = -n_4\mp\ii \cc\wt\nu,
\ede
where $n_i$ take nonnegative integers and $\cc=\pm$. Notice that since the two massive scalars in the loop are identical, only the combination $\cc_1=\cc_2=\cc_3=\cc_4=\cc$ in \eqref{eq_Specpole} can generate the nonanalyticity in $k_s$, and the residue sum is:
\begin{align}
\mathcal I_\text{NS}(r_1,r_2) =&~ \FR{r_1^4r_2^4}{128\pi^{7/2}} \sum_{\cc=\pm}
(1 -\cosh 2\pi\wt\nu)\Big(\FR{r_1r_2}4\Big)^{2\cc\ii\wt\nu}\n\\
&\times \sum_{n_1,\cdots,n_4=0}^\infty \FR{(-1)^{n_{1234}}}{n_1!n_2!n_3!n_4!}\Big(\FR{r_1}2\Big)^{2n_{13}}\Big(\FR{r_2}2\Big)^{2n_{24}}\prod_{i=1}^4\Big(\FR32+2n_i+\cc\ii\wt\nu\Big)\n\\&\times\Gamma\Big[ 4+2n_{13}+2\cc\ii\wt\nu,4+2n_{24}+2\cc\ii\wt\nu\Big]\n\\
& \times 
\Gamma\left[\bgm
-n_{1234}-2\cc\ii\wt\nu-\fr32,\fr32+n_{12}+\cc\ii\wt\nu,\fr32+n_{34}+\cc\ii\wt\nu\\ 3+n_{1234}+2\cc\ii\wt\nu,-n_{12}-\cc\ii\wt\nu,-n_{34}-\cc\ii\wt\nu
\edm\right]\n\\
&\times\Gamma\Big[ -n_1 -\cc\ii\wt\nu, -n_2 -\cc\ii\wt\nu, -n_3 -\cc\ii\wt\nu, -n_4 -\cc\ii\wt\nu\Big].
\end{align}
\paragraph{Local signal}
The local signal requires taking the loop UV pole \eqref{eq_UVpole} from the loop integral $\mathcal L$, together with spectrum poles for $s_1$ and $s_3$:
\bge
s_1 = -n_1-\cc\FR{\ii\wt\nu}2,\qquad s_3=-n_3-\cc\FR{\ii\wt\nu}2,\qquad s_4 = \FR32-m-s_{123} = \FR32 + m + n_{13} +\cc\ii\wt\nu - s_2,
\ede
where $n_{1,3}$ and $m$ take values in nonnegative integers, and $\cc=\pm$. Similarly, only spectral poles \eqref{eq_Specpole} with $\cc_1=\cc_3=\cc$ can give rise to the local signal. Summing up the residues, we obtain:
\begin{align}
\label{eq_LSintermediate}
\mathcal I_{\text{LS},>}(r_1,r_2)= &~ \FR{r_1^4r_2}{16\pi^{7/2}} \sum_{\cc=\pm}(1-\cosh 2\pi\wt\nu) \Big(\FR{r_1}{r_2}\Big)^{2\cc\ii\wt\nu}
\sum_{m,n_1,n_3=0}^\infty \FR{(-1)^{m+n_{13}}}{m!n_1!n_3!}
\Big(\FR{r_2}2\Big)^{2m}\Big(\FR{r_1}{r_2}\Big)^{2n_{13}}\n\\
&\times
\Gamma\Big[4+2n_{13}+2\cc\ii\wt\nu,1+2m-2n_{13}-2\cc\ii\wt\nu\Big]\n\\
&\times h_m\Big(-n_1-\cc\FR{\ii\wt\nu}2,-n_3-\cc\FR{\ii\wt\nu}2\Big) \times \Gamma\left[\bgm -n_1-\cc\ii\wt\nu,-n_3-\cc\ii\wt\nu\\ \fr32+m \edm\right],
\end{align}
where the function $h_m$ is the remaining single-layer Mellin integral over $s_2$:
\begin{align}
\label{eq_Hm}
h_m(s_1,s_3) \equiv &~\Big(\FR32-s_1\Big)\Big(\FR32-s_3\Big)\int_{-\ii\infty}^{+\ii\infty} \FR{\di s_2}{2\pi\ii}\,
\Big(\FR32-s_2\Big)(m+s_{123}\Big)
(s_{12})_m\Big(\FR32-m-s_{12}\Big)_m\n\\
&\times \Gamma\Big[
s_2+\cc\FR{\ii\wt\nu}2,s_2-\cc\FR{\ii\wt\nu}2, \FR32-m-s_{123} + \cc\FR{\ii\wt\nu}2,\FR32-m-s_{123} - \cc\FR{\ii\wt\nu}2
 \Big].
\end{align}
We can further finish this integral using Barnes' lemma \eqref{eq_Barnes}, similar to the calculation of function $f_m$ \eqref{eq_Fm}.
We put the calculation of $h_m$ in App.\ \ref{app_hm}, and list its expression in \eqref{eq_hm}. We plug \eqref{eq_hm} into \eqref{eq_LSintermediate} and obtain the final result for the local signal:
\begin{align}
\mathcal I_{\text{LS},>}(r_1,r_2)= &~ \FR{r_1^4r_2}{16\pi^{7/2}} \sum_{\cc=\pm}(1-\cosh 2\pi\wt\nu) \Big(\FR{r_1}{r_2}\Big)^{2\cc\ii\wt\nu}
\sum_{m,n_1,n_3=0}^\infty \FR{(-1)^{m+n_{13}}}{m!n_1!n_3!}
\Big(\FR{r_2}2\Big)^{2m}\Big(\FR{r_1}{r_2}\Big)^{2n_{13}}\n\\
&\times \Gamma\left[\bgm -n_1-\cc\ii\wt\nu,-n_3-\cc\ii\wt\nu\\ \fr32+m \edm\right]\times
\Gamma\Big[4+2n_{13}+2\cc\ii\wt\nu,1+2m-2n_{13}-2\cc\ii\wt\nu\Big]\n\\
&\times \Big(\FR32+n_1+\FR{\cc\ii\wt\nu}2\Big)\Big(\FR32+n_3+\FR{\cc\ii\wt\nu}2\Big)\sum_{t_1,t_3=0}^m (-1)^{t_{13}}\binom m{t_1}\binom m{t_3}\n\\
&\times
(1+n_1 -t_1)_{t_1} (1+n_3 -t_3)_{t_3}\times \Gamma\Big(\FR32+m+n_{13}-t_{13}\Big)\n\\
&\times \bigg\{
\Gamma\left[
\bgm
\fr72-m+n_{13}+2\cc\ii\wt\nu,\fr52+n_{13}+\cc\ii\wt\nu-t_1,\fr52+n_{13}+\cc\ii\wt\nu-t_3\\
5+2n_{13}+2\cc\ii\wt\nu
\edm
\right]
\n\\
&+\FR{3+\cc\ii\wt\nu}2\Big(m - n_{13}-\FR{3\cc\ii\wt\nu}{2}\Big)
\n\\
&\times \Gamma\left[\bgm
\fr32-m+n_{13}+2\cc\ii\wt\nu,\fr32+n_{13}+\cc\ii\wt\nu-t_1,\fr32+n_{13}+\cc\ii\wt\nu-t_3\\
3+2n_{13}+2\cc\ii\wt\nu
\edm
\right]\bigg\}.
\end{align}

\paragraph{Hierarchical squeezed limit}
Similar to the covariant case \eqref{eq_massless4pt}, the signals dominate in the hierarchical limit $r_1\ll r_2\ll 1$, where the expressions for both nonlocal and local signals can be greatly simplified:
\begin{align}
\lim_{r_1\ll r_2\ll 1} \mathcal I_\text{NS}(r_1,r_2) =&~ \FR{r_1^4r_2^4}{128\pi^{7/2}} \sum_{\cc=\pm}
(1 -\cosh 2\pi\wt\nu)\Big(\FR{r_1r_2}4\Big)^{2\cc\ii\wt\nu}\Big(\FR32+\cc\ii\wt\nu\Big)^4\n\\
&\times\Gamma\Big[ 4+2\cc\ii\wt\nu,4+2\cc\ii\wt\nu,-\cc\ii\wt\nu,-\cc\ii\wt\nu\Big]\n\\
& \times 
\Gamma\left[\bgm
-\fr32-2\cc\ii\wt\nu,\fr32+\cc\ii\wt\nu,\fr32+\cc\ii\wt\nu
\\ 3+2\cc\ii\wt\nu
\edm\right],\\
\lim_{r_1\ll r_2\ll 1} \mathcal I_\text{LS}(r_1,r_2) =&~ \FR{r_1^4r_2}{16\pi^{7/2}} \sum_{\cc=\pm}
(1 -\cosh 2\pi\wt\nu)\Big(\FR{r_1}{r_2}\Big)^{2\cc\ii\wt\nu}\Big(\FR32+\cc\ii\wt\nu\Big)^2\n\\
&\times \Big[\FR{(\fr32+2\cc\ii\wt\nu)_2}{(3+2\cc\ii\wt\nu)_2}\Big(\FR32+\cc\ii\wt\nu\Big)^2 - \FR{3\cc\ii\wt\nu}2 \FR{3+\cc\ii\wt\nu}2 \Big]\n\\
&\times\Gamma\Big[ 4+2\cc\ii\wt\nu,1-2\cc\ii\wt\nu,-\cc\ii\wt\nu,-\cc\ii\wt\nu\Big]\n\\
& \times \Gamma\left[\bgm
\fr32+2\cc\ii\wt\nu,\fr32+\cc\ii\wt\nu,\fr32+\cc\ii\wt\nu
\\ 3+2\cc\ii\wt\nu
\edm\right].
\end{align}

We find both the nonlocal and local signals still suffer from Boltzmann suppression with the factor $e^{-2\pi\wt\nu}$, but have an extra factor $\wt\nu^4 \sim m^4$ in the large mass limit compared to the signal part of the corresponding covariant loop seed integral, namely \eqref{eq_NSlimit} and \eqref{eq_LSlimit} with $\wt\nu_1=\wt\nu_2=\wt\nu$.
However, this does not mean the signal generated from interaction \eqref{eq_boostbreakingL} is larger than the covariant case, since the coupling in the dS-boost-breaking interaction should be more suppressed by $1/\Lambda^2$, where $\Lambda>m$ is the UV cutoff, because of the extra two derivatives on $\si$. Therefore, the signals from interaction \eqref{eq_boostbreakingL} should be as order $\mathcal O(m^4/\Lambda^4)$ in the large mass limit compared to the signals generated from the covariant bubble. 
Nevertheless, there are other models with nontrivial dS-boost breaking that can produce notably larger signals than that from the covariant bubble, where we can use the same method to derive their analytical results. We will leave this to future study.

\begin{figure}
\centering 
\includegraphics[width=0.45\textwidth]{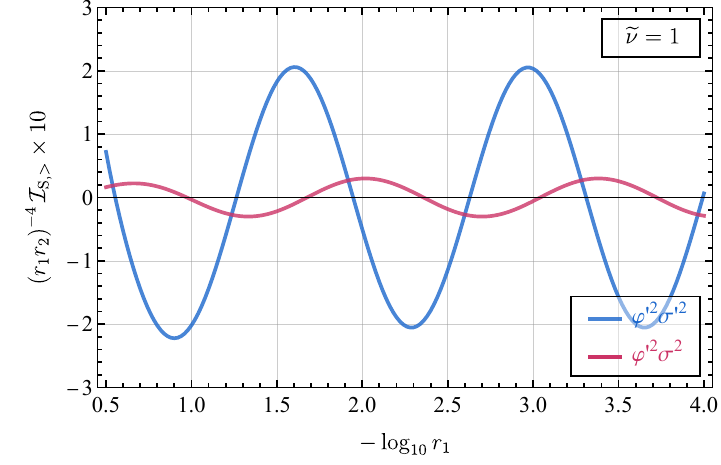}\qquad 
\includegraphics[width=0.45\textwidth]{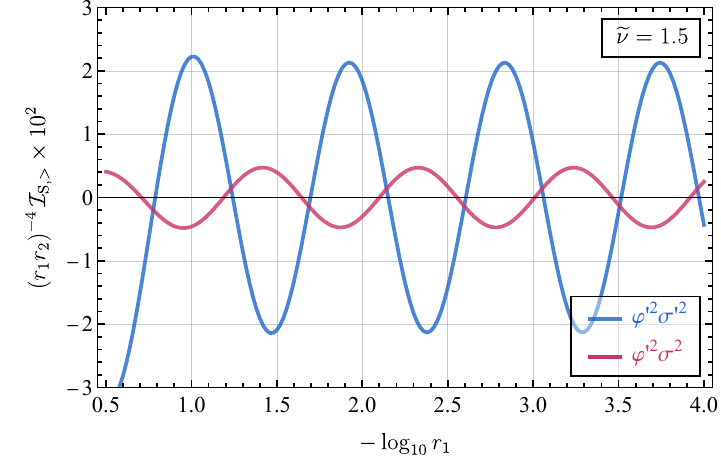}
\caption{
The signals of the 1-loop inflaton trispectra mediated by a pair of heavy scalars with mass $\wt\nu$, with or without dS-boost breaking.
The magenta curve shows the signal from the dS-boost-breaking interaction \eqref{eq_boostbreakingL}, and the blue curve shows the corresponding signal from a covariant bubble \eqref{eq_massless4pt}.
Here we fix $r_2 = 0.7$ and vary $r_1 \in (10^{-4},10^{-1/2})$.}
  \label{fig_boostbreaking}
\end{figure}

We plot the full signal part $\mathcal I_{\text{S},>} \equiv \mathcal I_{\text{NS}}+\mathcal I_{\text{LS},>}$ in Fig.\ \ref{fig_boostbreaking} and compare them with signals from the same bubble but with a covariant interaction. As we can see, the signals from boost-breaking and covariant bubbles have the same frequency $\omega = 2\wt\nu$, but with different phases (and also amplitudes), which in principle can help to break the degeneracy.

\section{Conclusions and Outlooks}
\label{sec_conclusion}
Loops are crucial but extremely challenging, especially in curved spacetime. Most methods developed for tree-level calculations in dS space do not apply at the loop level. Consequently, our understanding of general cosmological correlators at the loop level is quite limited, particularly when dS boosts are naturally broken by the rolling of the inflaton background. These cases are especially intriguing from a phenomenological perspective, as they can generate significant cosmological collider signals for future cosmic observations.

On the other hand, the method of PMB representation we proposed in \cite{Qin:2022lva,Qin:2022fbv} is powerful at the tree level. Utilizing the dilation symmetry without any requirements on dS boosts, PMB representation enables us to complete the calculation for general tree graphs \cite{Qin:2022fbv,Xianyu:2023ytd} with arbitrary interactions, including adding chemical potentials \cite{Qin:2022lva,Qin:2022fbv} that break both P and CP.

In this work, we go beyond the tree level and apply PMB representation to cosmological correlators at the loop level. We use PMB representation to resolve the complicated mode functions in curved spacetime into power functions and finish the time and loop momentum integrals. We then evaluate the remaining multi-layer Mellin integrals using the residue theorem, collecting residues at the appropriate poles, aided by Barnes' lemma \eqref{eq_Barnes}.

As our first example and also a consistency check, we first consider the four-point cosmological correlator generated from covariant bubbles. The signal of the correlator arises solely from the factorized part, consistent with the cutting rules, and can be expressed as a convergent series free from UV divergence. In contrast, the background piece can receive contributions from both the factorized and the time-ordered parts, and can be represented as a single-layer divergent Mellin integral. We explicitly extract the divergence of the background piece and demonstrate that it can be canceled by introducing the appropriate counterterm, ensuring that the renormalization scheme is the same as in flat spacetime.

Since the PMB representation does not rely on dS boosts, we apply this method to a bubble diagram with nontrivial dS-boost breaking. The pole structure of the Mellin integrand is similar to that of the covariant case, allowing us to conduct a comparable calculation. We present the analytical expressions for the full signals along with the leading signals in the hierarchical squeezed limit, where the results are further simplified.

Building on this work, there are many directions to explore in the future. We will outline some of them below and hope to address them in subsequent research.

First, as mentioned in the Introduction, one of the greatest advantages of the PMB representation is that it relies solely on dilation symmetry in the bulk. Therefore, we believe the PMB representation can aid in computing cosmological correlators in bubble topologies with arbitrary interactions or field content. For instance, we only consider bubbles with zero angular momentum, but the method should apply to bubbles with tensor structures as well. It should also apply to bubbles composed of spinning particles, including fermions, whose spectral decomposition has not yet been thoroughly studied. We further believe this method remains valid when chemical potentials that break dS boosts are introduced for the heavy fields in the bubble, as we have seen in the toy model in Sec.\ \ref{sec_break}. With the complete results, we will be able to construct accurate templates for these models, which are of significant interest in phenomenology.

Second, it will be important to compare the results for covariant bubbles derived from the PMB representation with those from spectral decomposition. While we can demonstrate that our results match the known results from spectral decomposition \cite{Xianyu:2022jwk} order by order in the squeezed limit or numerically, the latter expressions have a more compact form, facilitating the study of the analytical structure of the correlators. If we can establish a connection between these two equivalent expressions, we could derive spectral decomposition purely from a bulk calculation. Additionally, we hope to rewrite the correlators of dS-boost-breaking bubbles in a more compact form to explore the general analytical structure at the loop level, and to study the effect of dS boosts on the analytical behaviors. It will also help us to take the folded limit from the four-point functions to obtain bispectra and power spectra as in \cite{Qin:2023ejc,Xianyu:2022jwk}.

Third, the PMB representation can assist us in exploring loop-level correlators with more complex topologies, such as triangles and boxes, which are also important channels in particle physics and phenomenology. Although we cannot express the analytical results of the time and loop integrals in Mellin space in terms of well-known functions for these diagrams, we can analyze the pole structure of the Mellin integrand following the spirit of Landau analysis \cite{Qin:Local}. Therefore, we hope to evaluate (some of) the Mellin integrals either analytically or numerically. Furthermore, if we obtain the (leading) signal that corresponds to the (leading) discontinuity of the correlator in the complex plane, we may be able to reconstruct the full correlator using the dispersion integral, as outlined in \cite{Liu:2024xyi}.
We leave all these open questions for future work.

\paragraph{Acknowledgments.} We thank Shuntaro Aoki, Daniel Green, Kshitij Gupta, Enrico Pajer, Xi Tong, Dong-Gang Wang, Zhong-Zhi Xianyu, Hongyu Zhang, and Yuhang Zhu for their helpful discussions. We also thank Shuntaro Aoki, Zhong-Zhi Xianyu, and Yuhang Zhu for their useful comments on the draft.
This work is supported by NSFC under Grant No.\ 12275146, the National Key R\&D Program of China (2021YFC2203100), and the Dushi Program of Tsinghua University.

\newpage
\begin{appendix}
\section{Useful Functions}
\label{app_A}
In this work we heavily use the following shorthand for the products and/or the fractions of Euler $\Gamma$ functions:
\begin{align}
&\Gamma\Big[a,\cdots,b\Big] \equiv \Gamma(a)\cdots \Gamma(b),\\
&\Gamma\left[\bgm a,\cdots,b\\ c,\cdots,d\edm\right] \equiv \FR{\Gamma(a)\cdots \Gamma(b)}{\Gamma(c)\cdots \Gamma(d)}.
\end{align}

We also use a lot of the (generalized) hypergeometric function, which is defined by:
\bge
{}_p\mathrm F_q\left[\bgm a_1,\cdots,a_p\\b_1,\cdots,b_q\edm\middle|z\right] \equiv \sum_{n=0}^\infty \FR{(a_1)_n\cdots(a_p)_n}{(b_1)_n\cdots(b_q)_n}\FR{z^n}{n!}.
\ede
We will only encounter the case with $q=p+1$, where the above series expansion is always convergent within the disk $|z|<1$, and the (generalized) hypergeometric function is defined via analytic continuation on the complex plane.

For convenience we also define the regularized (generalized) hypergeometric function:
\bge
{}_p\wh {\mathrm F}_q\left[\bgm a_1,\cdots,a_p\\b_1,\cdots,b_q\edm\middle|z\right] \equiv \FR{1}{\Gamma[b_1,\cdots,b_q]}{}_p\mathrm F_q\left[\bgm a_1,\cdots,a_p\\b_1,\cdots,b_q\edm\middle|z\right],
\ede
which is an entire function of all parameters $(a_1,\cdots,a_p,b_1,\cdots b_q)$ at all regular points of $z$.

Finally, we define the dressed (generalized) hypergeometric function:
\bge
{}_p\mathcal F_q\left[\bgm a_1,\cdots,a_p\\b_1,\cdots,b_q\edm\middle|z\right] \equiv \Gamma\Big[a_1,\cdots,a_p \Big]{}_p\wh{\mathrm F}_q\left[\bgm a_1,\cdots,a_p\\b_1,\cdots,b_q\edm\middle|z\right].
\ede
Therefore, the pole structure of the dressed (generalized) hypergeometric function as a function of parameters (at regular points of $z$) is fully encoded in the product of $\Gamma$ functions $\Gamma[a_1,\cdots,a_p]$, namely the poles are:
\bge
a_i = -n,\qquad i=1,\cdots,p,\quad n=0,1,\cdots,
\ede
and the residues can be easily read out.

The asymptotic behaviors of these special functions are particularly useful. For the $\Gamma$ function, we have Stirling’s formula \cite{nist:dlmf}:
\begin{equation}
\label{eq_GammaAsym}
\lim_{z\to\infty} \Gamma(az+b) \sim \sqrt{2\pi} e^{-az}(az)^{az+b-1/2},
\end{equation}
provided that $z\to\infty$ in the sector $|\text{Arg}\, z| \leq \pi - \de$.
And for the hypergeometric, we are concerned with the following special limit (at a regular point of $z$):
\bge
\label{eq_HyperAsym}
\lim_{x\to\infty} {}_2\mathrm F_1\left[\bgm a, b+x \\ c+x \edm\middle| z \right] = \lim_{x\to\infty} \sum_{n=0}^\infty \FR{(a)_n(b+x)_n}{(c+x)_n}\FR{z^n}{n!}=\sum_{n=0}^\infty\FR{(a)_nz^n}{n!} = (1-z)^a.
\ede

\section{Useful Integrals}
\label{app_B}

\paragraph{Loop integral}
The loop integral \eqref{eq_L} for the bubble diagram we encounter can be easily carried out:
\bge
 \int \FR{\di^3\mb q}{(2\pi)^3}\,k_s^{-2s_{12}}|\mb k_s + \mb q|^{-2s_{34}}=\FR{k_s^{3-2s_{1234}}}{(4\pi)^{3/2}}\Gamma\left[\bgm
s_{1234}-\fr32,\fr32-s_{12},\fr32-s_{34}\\ 3-s_{1234},s_{12},s_{34}
\edm\right].
 \ede
 
 \paragraph{Time integrals}
We also encounter the single-site time integral and two-site time-ordered integrals, which are:
\begin{align}
\label{eq_1site}
&\int_{-\infty}^0 \di\tau\,(-\tau)^{p-1}e^{\pm\ii E\tau} = e^{\mp\ii p\pi/2}E^{-p}\Gamma(p),\\
\label{eq_2site}
&\int_{-\infty}^0 \di\tau_1\di\tau_2\,\theta(\tau_2-\tau_1)(-\tau_1)^{p_1-1}(-\tau_2)^{p_2-1}e^{\pm\ii (E_1\tau_1+E_2\tau_2)} = e^{\mp\ii p_{12}\pi/2}E_1^{-p_{12}}{}_2\mathcal F_1\left[\bgm p_2,p_{12}\\1+p_2 \edm\middle|-\FR{E_2}{E_1}\right].
\end{align}

Using \eqref{eq_1site} and \eqref{eq_2site}, together with the decomposition of the product of nesting functions \eqref{eq_Nsimp}, we can derive the time integrals \eqref{eq_T} appearing in the loop seed integral \eqref{eq_loopseed}:
\begin{align}
\mathcal T(k_{12},k_{34}) =&~ \sum_{\aa,\bb=\pm}(-\aa\bb) \int_{-\infty}^0 \di\tau_1\di\tau_2\,(-\tau_1)^{3+p_1-2s_{13}}(-\tau_2)^{3+p_2-2s_{24}}e^{\aa\ii k_{12}\tau_1+\bb\ii k_{34}\tau_2}\n\\
&\times N_{\aa\bb}(s_1,s_2;\tau_1,\tau_2)N_{\aa\bb}(s_3,s_4;\tau_1,\tau_2)\n\\
=&~ \sum_{\aa,\bb=\pm}(-\aa\bb) \int_{-\infty}^0 \di\tau_1\di\tau_2\,(-\tau_1)^{3+p_1-2s_{13}}(-\tau_2)^{3+p_2-2s_{24}}e^{\aa\ii k_{12}\tau_1+\bb\ii k_{34}\tau_2}\n\\
&\times N_{-\bb,\bb}(s_1,s_2;\tau_1,\tau_2)N_{-\bb,\bb}(s_3,s_4;\tau_1,\tau_2)\n\\
&-\sum_{\bb=\pm}\int_{-\infty}^0 \di\tau_1\di\tau_2\,\theta(\tau_2-\tau_1)(-\tau_1)^{3+p_1-2s_{13}}(-\tau_2)^{3+p_2-2s_{24}}e^{\bb\ii (k_{12}\tau_1+k_{34}\tau_2)}\n\\
&\times  \Big[ N_{\bb,-\bb}(s_1,s_2)N_{\bb,-\bb}(s_3,s_4)-N_{-\bb,\bb}(s_1,s_2)N_{-\bb,\bb}(s_3,s_4)\Big]\n\\
=&~\FR{2^{2s_{1234}}}{8\pi^2}\bigg[\cos \FR{\pi \bar p_{12}}2 -\cos\Big(2\pi s_{13} - \FR{\pi p_{12}}2\Big)\bigg] k_{12}^{-4-p_1+2s_{13}} k_{34}^{-4-p_2+2s_{24}}\n\\&\times \Gamma\Big[ 4+p_1-2s_{13},4+p_2-2s_{24}\Big]\n\\
&+\FR{2^{2s_{1234}}}{4\pi^2} \sin[\pi(s_{13}-s_{24})]\sin\Big( \FR{\pi p_{12}}2 - \pi s_{1234}\Big) k_{12}^{-8-p_{12}+2s_{1234}}\n\\&\times {}_2\mathcal F_1\left[\bgm4+p_2-2s_{24},8+p_{12}-2s_{1234}\\5+p_2-2s_{24}\edm\middle| -\FR{k_{34}}{k_{12}}\right],
\end{align}
which gives the results \eqref{eq_Tresult}-\eqref{eq_TTO}.

\paragraph{Barnes' lemma}
Barnes' lemma \cite{bailey1935generalized} is the key tool when we compute the Mellin integrals at loop levels:
\bge
\label{eq_Barnes}
\int_{-\ii\infty}^{+\ii\infty} \FR{\di s}{2\pi\ii}\, \Gamma\Big[a+s,b+s,c-s,d-s\Big] = \Gamma\left[\bgm a+c,a+d,b+c,b+d\\a+b+c+d\edm\right],
\ede
provided that the contour in the complex plane is curved such that it separates the left and the right poles.
\section{Intermediate Steps}
\label{app_C}
\subsection{Calculation of $f_m$}
\label{app_fm}
In this appendix we calculate the integral $f_m(s_1,s_3)$ \eqref{eq_Fm}:
\begin{align}
f_m(s_1,s_3) \equiv &\int_{-\ii\infty}^{+\ii\infty} \FR{\di s_2}{2\pi\ii}\,(s_{12})_m\Big(\FR32-m-s_{12}\Big)_m\n\\
&\times \Gamma\Big[
s_2+\cc_1\FR{\ii\wt\nu_1}2,s_2-\cc_1\FR{\ii\wt\nu_1}2, \FR32-m-s_{123} + \cc_3\FR{\ii\wt\nu_2}2,\FR32-m-s_{123} - \cc_3\FR{\ii\wt\nu_2}2
 \Big],
\end{align}
where we have assigned two signs $\cc_{1,3}=\pm$ for later convenience.
The key observation is that the pochhammer functions can be absorbed into the $\Gamma$ functions using the identity:
\bge
\label{eq_identity}
(x+a)_m\Gamma(x) = \sum_{t=0}^{m}(-1)^t \binom mt (1-a-t)_t\Gamma(x+m-t).
\ede
In particular, we have:
\begin{align}
(s_{12})_m\Gamma\Big(s_2-\cc_1\FR{\ii\wt\nu_1}2\Big) =&~ \sum_{t_1=0}^m (-1)^{t_1}\binom m{t_1} \Big(1- s_1 -\cc_1\FR{\ii\wt\nu_1}2 -t_1 \Big)_{t_1}\n\\
&\times \Gamma\Big(s_2-\cc_1\FR{\ii\wt\nu_1}2 + m - t_1\Big),\\
\Big(\FR32-m-s_{12}\Big)_m \Gamma\Big(\FR32-m-s_{123}-\cc_3\FR{\ii\wt\nu_2}2\Big)
 =&~ \sum_{t_3=0}^m(-1)^{t_3}\binom m{t_3} \Big(1-s_3-\cc_3\FR{\ii\wt\nu_2}2-t_3\Big)_{t_3}\n\\
 &\times \Gamma\Big(\FR32-s_{123}-\cc_3\FR{\ii\wt\nu_2}2 - t_3\Big).
\end{align}
As a result, the integrand of $f_m$ becomes the standard form of the Barnes' lemma \eqref{eq_Barnes}:
\begin{align}
f_m(s_1,s_3) =&~  \sum_{t_1,t_3=0}^m (-1)^{t_{13}}\binom m{t_1}\binom m{t_3}\times
\Big(1- s_1 -\cc_1\FR{\ii\wt\nu_1}2 -t_1\Big)_{t_1}\Big(1- s_3 -\cc_3\FR{\ii\wt\nu_2}2 -t_3\Big)_{t_3}
\n\\
&\times \int_{-\ii\infty}^{+\ii\infty} \FR{\di s_2}{2\pi\ii}\, \Gamma\Big[
s_2+\cc_1\FR{\ii\wt\nu_1}2,s_2-\cc_1\FR{\ii\wt\nu_1}2+m-t_1\Big]\n\\
&\times \Gamma\Big[\FR32-m-s_{123}+\cc_3\FR{\ii\wt\nu_2}2,\FR32-s_{123}-\cc_3\FR{\ii\wt\nu_2}2-t_3
\Big],
\end{align}
and thus we can compute the integral directly:
\begin{align}
f_m(s_1,s_3) =&~  \sum_{t_1,t_3=0}^m (-1)^{t_{13}}\binom m{t_1}\binom m{t_3}\times
\FR{(1- s_1 -\cc_1\fr{\ii\wt\nu_1}2 -t_1)_{t_1}(1- s_3 -\cc_3\fr{\ii\wt\nu_2}2 -t_3)_{t_3}}{\Gamma(3-2s_{13}-t_{13})}
\n\\
&\times \Gamma\Big[ \FR32-m -s_{13} + \FR{\cc_1\ii\wt\nu_1+\cc_3\ii\wt\nu_2}{2},
\FR32+m-s_{13}- \FR{\cc_1\ii\wt\nu_1+\cc_3\ii\wt\nu_2}{2} -t_{13}\Big]\n\\
&\times \Gamma\Big[
 \FR32 -s_{13} + \FR{\cc_1\ii\wt\nu_1-\cc_3\ii\wt\nu_2}{2}-t_3,
\FR32-s_{13}- \FR{\cc_1\ii\wt\nu_1-\cc_3\ii\wt\nu_2}{2} -t_1
\Big].
\end{align}

\subsection{Calculation of $g_m$}
\label{app_gm}
In this appendix we calculate the integral $g_m(S)$ \eqref{eq_Gm}:
\begin{align}
\label{eq_step0}
g_m(S) 
=&~ \sum_{t_1,t_3=0}^m (-1)^{t_{13}}\binom m{t_1}\binom m{t_3}\times \FR1{\Gamma(3-2S-t_{13})}\n\\
&\times \Gamma\Big[ \FR32-m -S + \FR{\ii\wt\nu_1+\ii\wt\nu_2}{2},
\FR32+m-S- \FR{\ii\wt\nu_1+\ii\wt\nu_2}{2} -t_{13}\Big]\n\\
&\times \Gamma\Big[
 \FR32 -S + \FR{\ii\wt\nu_1-\ii\wt\nu_2}{2}-t_3,
\FR32-S- \FR{\ii\wt\nu_1-\ii\wt\nu_2}{2} -t_1
\Big]\n\\
&\times \int_{-\ii\infty}^{+\ii\infty}\FR{\di s_1}{2\pi\ii}\,
\Big(1- s_1 -\FR{\ii\wt\nu_1}2 -t_1\Big)_{t_1}\Big(1- S+s_1 -\FR{\ii\wt\nu_2}2 -t_3\Big)_{t_3}
\n\\
&\times\Gamma\Big[ s_1+\FR{\ii\wt\nu_1}2,s_1-\FR{\ii\wt\nu_1}2,S-s_1+\FR{\ii\wt\nu_2}2, S-s_1-\FR{\ii\wt\nu_2}2\Big],
\end{align}
where we have inserted the expression for the function $f_m$ \eqref{eq_fm} with the fixed signs $\cc_1=\cc_3=+$.

Similar to the calculation of $f_m$ in App.\ \ref{app_fm}, we can use the Barnes' lemma \eqref{eq_Barnes} and the identity \eqref{eq_identity} to compute the integral over $s_1$. First use \eqref{eq_identity} to rewrite the integrand:
\begin{align}
\label{eq_step1}
I\equiv & \int_{-\ii\infty}^{+\ii\infty}\FR{\di s_1}{2\pi\ii}\,
\Big(1- s_1 -\FR{\ii\wt\nu_1}2 -t_1\Big)_{t_1}\Big(1- S + s_1 -\FR{\ii\wt\nu_2}2 -t_3\Big)_{t_3}
\n\\
&\times\Gamma\Big[ s_1+\FR{\ii\wt\nu_1}2,s_1-\FR{\ii\wt\nu_1}2,S-s_1+\FR{\ii\wt\nu_2}2, S-s_1-\FR{\ii\wt\nu_2}2\Big]\n\\
=&~\sum_{\ell_1=0}^{t_1}\sum_{\ell_3=0}^{t_3}(-1)^{\ell_{13}}\binom{t_1}{\ell_1}\binom{t_3}{\ell_3}\Big(S+t_1 + \FR{\ii\wt\nu_1-\ii\wt\nu_2}{2}-\ell_1\Big)_{\ell_1}
\Big(S+t_3 - \FR{\ii\wt\nu_1-\ii\wt\nu_2}{2}-\ell_3\Big)_{\ell_3}\n\\
&\times  \int_{-\ii\infty}^{+\ii\infty}\FR{\di s_1}{2\pi\ii}\,\Gamma\Big[
s_1+\FR{\ii\wt\nu_1}2, s_1-\FR{\ii\wt\nu_1}2+t_3-\ell_3, S-s_1+\FR{\ii\wt\nu_2}2,
S -s_1 - \FR{\ii\wt\nu_2}{2} + t_1-\ell_1\Big],
\end{align}
then the integral of $s_1$ on the last line of \eqref{eq_step1} can be completed via Barnes' lemma \eqref{eq_Barnes}:
\begin{align}
\label{eq_step2}
&\int_{-\ii\infty}^{+\ii\infty}\FR{\di s_1}{2\pi\ii}\,\Gamma\Big[
s_1+\FR{\ii\wt\nu_1}2, s_1-\FR{\ii\wt\nu_1}2+t_3-\ell_3, S-s_1+\FR{\ii\wt\nu_2}2,
S -s_1 - \FR{\ii\wt\nu_2}{2} + t_1-\ell_1\Big]\n\\
=&~\Gamma\Big[S+\FR{\ii\wt\nu_1-\ii\wt\nu_2}2+t_1-\ell_1,S-\FR{\ii\wt\nu_1-\ii\wt\nu_2}2+t_3-\ell_3\Big]\n\\
&\times
\Gamma\left[\bgm 
S+\FR{\ii\wt\nu_1+\ii\wt\nu_2}2,S-\FR{\ii\wt\nu_1+\ii\wt\nu_2}2+t_{13}-\ell_{13}
\\2S+t_{13}-\ell_{13}
\edm\right].
\end{align}
Plugging \eqref{eq_step2} into \eqref{eq_step1}, and then back into \eqref{eq_step0}, we can finish the calculation of $g_m(S)$, and the expression can be further simplified. Notice that the pochhammer functions on the second to last line in \eqref{eq_step1} can be absorbed into the $\Gamma$ functions in \eqref{eq_step2}, we find:
\begin{align}
I =&~\sum_{\ell_1=0}^{t_1}\sum_{\ell_3=0}^{t_3}(-1)^{\ell_{13}}\binom{t_1}{\ell_1}\binom{t_3}{\ell_3}\times \Gamma\Big[S+\FR{\ii\wt\nu_1-\ii\wt\nu_2}2+t_1,S-\FR{\ii\wt\nu_1-\ii\wt\nu_2}2+t_3\Big]\n\\
&\times \Gamma\left[\bgm 
S+\FR{\ii\wt\nu_1+\ii\wt\nu_2}2,S-\FR{\ii\wt\nu_1+\ii\wt\nu_2}2+t_{13}-\ell_{13}
\\2S+t_{13}-\ell_{13}
\edm\right],
\end{align}
and we can finish the summation of $\ell_1$ and $\ell_3$ to obtain:
\bge
\label{eq_step3}
I = (-1)^{t_{13}} \Gamma\left[\bgm 
S+\FR{\ii\wt\nu_1+\ii\wt\nu_2}2+t_{13},S-\FR{\ii\wt\nu_1+\ii\wt\nu_2}2,S+\FR{\ii\wt\nu_1-\ii\wt\nu_2}2+t_1,S-\FR{\ii\wt\nu_1-\ii\wt\nu_2}2+t_3
\\2S+t_{13}
\edm\right].
\ede
Plugging \eqref{eq_step3} into \eqref{eq_step0}, we finally obtain the expression for $g_m$:
\begin{align}
\label{eq_gm}
g_m(S) =&~ 
\Gamma\Big[ \FR32-m -S + \FR{\ii\wt\nu_1+\ii\wt\nu_2}{2},S-\FR{\ii\wt\nu_1+\ii\wt\nu_2}2\Big]
\sum_{t_1,t_3=0}^m \FR{\binom m{t_1}\binom m{t_3}}{\Gamma[2S+t_{13},3-2S-t_{13}]}\n\\
&\times \Gamma\Big[
 \FR32 -S + \FR{\ii\wt\nu_1-\ii\wt\nu_2}{2}-t_3,
\FR32-S- \FR{\ii\wt\nu_1-\ii\wt\nu_2}{2} -t_1,\FR32+m-S- \FR{\ii\wt\nu_1+\ii\wt\nu_2}{2} -t_{13}
\Big]\n\\
&\times
 \Gamma\left[\bgm 
S+\FR{\ii\wt\nu_1+\ii\wt\nu_2}2+t_{13},S+\FR{\ii\wt\nu_1-\ii\wt\nu_2}2+t_1,S-\FR{\ii\wt\nu_1-\ii\wt\nu_2}2+t_3
\edm\right].
\end{align}

We can further identify some useful analytical behaviors of the function $g_m(S)$ through the expression \eqref{eq_gm}:
\begin{itemize}
\item
First, notice the factor:
\bge
\FR1{\Gamma(2S+t_{13})} = \FR{(2S+t_{13})_{\Delta n}}{\Gamma(2S+2m)},\qquad \Delta n = 2m-t_{13}\in \mathbb N,
\ede
we immediately know that $g_m(S)$ has a series of zeros:
\bge
\label{eq_zeros}
S = -m -\FR n2,\qquad n=0,1,\cdots.
\ede
Similarly, it has another series of zeros from the factor $1/\Gamma(3-2S-t_{13})$:
\bge
\label{eq_zeros2}
S=  \FR32 + \FR n2 , \qquad n=0,1,\cdots.
\ede
\item
Second, let us consider the asymptotic behavior of $g_m(S)$ when $S\to\pm\ii\infty$.
Using the asymptotic behavior of $\Gamma$ function \eqref{eq_GammaAsym}, we obtain:
\bge
\lim_{z\to\pm\ii\infty} \Gamma\Big[a+z,b-z\Big] 
\sim 2\pi e^{\pm\ii\pi(a-b)/2} e^{-\pi |z|} |z|^{a+b-1},
\ede
and plugging this into the expression of $g_m(S)$ \eqref{eq_gm}, we find:
\bge
\lim_{S\to\pm\ii\infty} g_m(S) \sim 2\ii\pi^3 e^{-2\pi |S|}\sum_{t_1,t_3=0}^m
\binom{m}{t_1}\binom{m}{t_3}(-1)^{t_{13}}.
\ede
Now notice that
\bge
\sum_{t_1,t_3=0}^m
\binom{m}{t_1}\binom{m}{t_3}(-1)^{t_{13}} = \begin{cases} 1&m=0,\\0&m=1,2,\cdots,\end{cases}
\ede
we finally obtain the asymptotic behavior of $g_m(S)$:
\bge
\label{eq_Slimit}
\lim_{S\to\pm\ii\infty} g_m(S) =\begin{cases}
 2\ii\pi^3 e^{-2\pi |S|} & m=0,\\
 \mathcal O\Big(\FR{e^{-2\pi |S|}}{|S|}\Big) & m=1,2,\cdots.
\end{cases}
\ede
\end{itemize}

\subsection{Calculation of $h_m$}
\label{app_hm}
In this appendix we calculate the integral $h_m(s_1,s_3)$ \eqref{eq_Hm}:
\begin{align}
h_m(s_1,s_3) = &~\Big(\FR32-s_1\Big)\Big(\FR32-s_3\Big)\int_{-\ii\infty}^{+\ii\infty} \FR{\di s_2}{2\pi\ii}\,
\Big(\FR32-s_2\Big)(m+s_{123}\Big)
(s_{12})_m\Big(\FR32-m-s_{12}\Big)_m\n\\
&\times \Gamma\Big[
s_2+\cc\FR{\ii\wt\nu}2,s_2-\cc\FR{\ii\wt\nu}2, \FR32-m-s_{123} + \cc\FR{\ii\wt\nu}2,\FR32-m-s_{123} - \cc\FR{\ii\wt\nu}2
 \Big],
\end{align}
Similar to the calculation of $f_m$ in App.\ \ref{app_fm}, we can first absorb the two pochhammer functions into the $\Gamma$ functions:
\begin{align}
h_m(s_1,s_3) \equiv &~\Big(\FR32-s_1\Big)\Big(\FR32-s_3\Big)\sum_{t_1,t_3=0}^m (-1)^{t_{13}}\binom m{t_1}\binom m{t_3}\n\\
&\times\Big(1- s_1 -\cc\FR{\ii\wt\nu_1}2 -t_1\Big)_{t_1} \Big(1- s_3 -\cc\FR{\ii\wt\nu_2}2 -t_3\Big)_{t_3}\n\\
&\times \int_{-\ii\infty}^{+\ii\infty} \FR{\di s_2}{2\pi\ii}\,
\Big(\FR32-s_2\Big)(m+s_{123}\Big)\Gamma\Big[s_2+\cc\FR{\ii\wt\nu}2,s_2-\cc\FR{\ii\wt\nu}2+m-t_1\Big]
\n\\
&\times \Gamma\Big[\FR32-m-s_{123}+\cc\FR{\ii\wt\nu}2,\FR32-s_{123}-\cc\FR{\ii\wt\nu}2-t_3
\Big].
\end{align}
The polynomial $(3/2-s_2)(m+s_{123})$ in the integrand can be written as:
\bge
(3/2-s_2)(m+s_{123}) = \Big(s_2+\cc\FR{\ii\wt\nu}2\Big) \Big(\FR32-m-s_{123}+\cc\FR{\ii\wt\nu}2\Big)
+ \FR{3+\cc\ii\wt\nu}2\Big(m+ s_{13}-\FR{\cc\ii\wt\nu}{2}\Big),
\ede
and thus we can rewrite the integrand as the standard form of Barnes' lemma again:
\begin{align}
h_m(s_1,s_3) = &~\Big(\FR32-s_1\Big)\Big(\FR32-s_3\Big)\sum_{t_1,t_3=0}^m (-1)^{t_{13}}\binom m{t_1}\binom m{t_3}\n\\
&\times\Big(1- s_1 -\cc\FR{\ii\wt\nu_1}2 -t_1\Big)_{t_1} \Big(1- s_3 -\cc\FR{\ii\wt\nu_2}2 -t_3\Big)_{t_3}\n\\
&\times \bigg\{ \int_{-\ii\infty}^{+\ii\infty} \FR{\di s_2}{2\pi\ii}\,
\Gamma\Big[1+s_2+\cc\FR{\ii\wt\nu}2,s_2-\cc\FR{\ii\wt\nu}2+m-t_1\Big]
\n\\
&\times \Gamma\Big[\FR52-m-s_{123}+\cc\FR{\ii\wt\nu}2,\FR32-s_{123}-\cc\FR{\ii\wt\nu}2-t_3
\Big]\n\\
&+\FR{3+\cc\ii\wt\nu}2\Big(m+ s_{13}-\FR{\cc\ii\wt\nu}{2}\Big)\int_{-\ii\infty}^{+\ii\infty} \FR{\di s_2}{2\pi\ii}\,
\Gamma\Big[s_2+\cc\FR{\ii\wt\nu}2,s_2-\cc\FR{\ii\wt\nu}2+m-t_1\Big]
\n\\
&\times \Gamma\Big[\FR32-m-s_{123}+\cc\FR{\ii\wt\nu}2,\FR32-s_{123}-\cc\FR{\ii\wt\nu}2-t_3
\Big]\bigg\}.
\end{align}
Finishing the integral over $s_2$, we finally obtain:
\begin{align}
\label{eq_hm}
h_m(s_1,s_3) = &~\Big(\FR32-s_1\Big)\Big(\FR32-s_3\Big)\sum_{t_1,t_3=0}^m (-1)^{t_{13}}\binom m{t_1}\binom m{t_3}\n\\
&\times\Big(1- s_1 -\cc\FR{\ii\wt\nu_1}2 -t_1\Big)_{t_1} \Big(1- s_3 -\cc\FR{\ii\wt\nu_2}2 -t_3\Big)_{t_3}\n\\
&\times \bigg\{
\Gamma\left[
\bgm
\fr72-m-s_{13}+\cc\ii\wt\nu,\fr32+m-s_{13}-\cc\ii\wt\nu-t_{13},\fr52-s_{13}-t_1,\fr52-s_{13}-t_3\\
5-2s_{13}
\edm
\right]
\n\\
&+\FR{3+\cc\ii\wt\nu}2\Big(m+ s_{13}-\FR{\cc\ii\wt\nu}{2}\Big)
\n\\
&\times \Gamma\left[\bgm
\fr32-m-s_{13}+\cc\ii\wt\nu,\fr32+m-s_{13}-\cc\ii\wt\nu-t_{13},\fr32-s_{13}-t_1,\fr32-s_{13}-t_3\\
3-2s_{13}
\edm
\right]\bigg\}.
\end{align}
\end{appendix}

\newpage
\bibliography{CosmoCollider} 
\bibliographystyle{utphys}

\end{document}